\documentclass[twocolumn]{aa}
\usepackage{graphicx}
\usepackage{txfonts}
\usepackage{hyperref}
\usepackage{natbib}
\usepackage{soul}
\usepackage{xcolor}
\begin{document}

    \title{A Self-consistent Model for the Generation of Coronal Condensations: The Effects of Turbulent Wave Dissipation}
    \titlerunning{UAWSoM}

\author{
  M. McMurdo\inst{1}\thanks{Corresponding author: \email{max.mcmurdo@kuleuven.be}} \and
  V. Jer\v{c}i\'c\inst{2} \and
  T. Van Doorsselaere\inst{1} \and 
  C. Froment\inst{3}}

\institute{
  Centre for mathematical Plasma Astrophysics, Department of Mathematics, KU Leuven, Celestijnenlaan 200B Bus 2400, 3001 Leuven, Belgium \and
  NASA Goddard Space Flight Center, Greenbelt, MD, USA \and
  LPC2E, OSUC, Univ Orleans, CNRS, CNES, F-45071 Orleans, France
  }
   \date{Received ; accepted }

  \abstract
   {Prominences and coronal rain are two manifestations of coronal condensations whose formation mechanisms remain only partially understood. While previous studies have explored how localised steady or stochastic heating can trigger thermal instability and condensations, the role of wave-driven energy transport and dissipation has not been extensively investigated. In the solar atmosphere, Alfv\'en and kink waves are an abundant source of energy capable of contributing both to coronal heating and plasma structuring, making them compelling candidates in the formation processes of the structures we observe in the corona.}
   {We aim to explore how Alfv\'en and kink wave energy and their dissipation influence the formation and evolution of condensations in coronal magnetic structures. In particular, we investigate each wave's role independently of parameterised external heating mechanisms to isolate whether either wave can trigger condensations.}
   {We perform 2.5D magnetohydrodynamic (MHD) simulations using the open-source \texttt{MPI-AMRVAC} code extended with the newly developed Uniturbulence and Alfv\'en Wave Solar Model (UAWSoM) module. This module self-consistently evolves Alfv\'en and kink wave energy and their feedback on the plasma through wave pressure and heating terms. We examine the conditions under which condensations form, their morphology, and their dynamical evolution.}
   {Our results show that for our range of parameters, in 2.5D MHD, Alfv\'en waves did not trigger the formation of prominences or condensations. We find that if the radius of the fine-scale structures supporting the kink waves is of the order of 100 km, then kink waves can trigger coronal rain that differs considerably in its morphology from coronal rain triggered by parameterised heating functions.}
   {Through the use of UAWSoM, we have established that Alfv\'en and kink waves leave distinct thermodynamic signatures on coronal loops, given the same wave energy injection. This investigation establishes a proof-of-concept that warrants further investigation into the distinct roles that Alfv\'en and kink wave heating can have on the formation of coronal condensations. Future, more complete models, have the potential to provide strong evidence for heating-mechanism-specific identification of observable properties such as prominence formation and coronal rain.}


\maketitle
\nolinenumbers
\section{Introduction}\label{sec:Introduction}

Solar prominences, filaments, and coronal rain are all manifestations of cool, dense plasma condensations embedded within the hot solar corona. Prominences and filaments refer to the same physical structures, appearing in emission above the solar limb and in absorption against the disk, respectively. Coronal rain likewise consists of cool condensations that form within coronal loops and subsequently drain toward the lower atmosphere. Although these phenomena are often treated separately, they are closely related and share many of the same underlying physical processes, including localised heating, thermal conduction, radiative losses, and the onset of thermal instability \citep{Field1965, Mackay2010, Kaneko2015, Kaneko2017, Xia2011, Zhao2022, Brughmans2022}. Their plasma is typically two orders of magnitude cooler and denser than the surrounding coronal environment, and their morphology and dynamics exhibit considerable diversity. Observations reveal prominences composed of both vertically oriented, Rayleigh-Taylor-unstable columns and horizontally aligned, thread-like structures \citep{Berger2010, Parenti2014, Hillier2021}. These features are highly dynamic, often exhibiting fine-scale flows, oscillations, and partial eruptions, indicating that prominences rarely attain strict thermal or mechanical equilibrium.

Condensations can form when plasma is heated in a spatially non-uniform manner that destabilises the loop and leads to in situ cooling and condensation \citep{Mackay2010, Kaneko2015, Kaneko2017, Xia2011, Brughmans2022, Jenkins2021, Jenkins2022, Zhao2022}. This process is commonly interpreted in terms of thermal instability \citep{Field1965, Donne2024, Keppens2025, Kelly2026}, which is triggered when heating, radiative losses, and conductive transport become imbalanced and is specifically related to the evaporation-condensation mechanism, known as thermal non-equilibrium \citep{Xia2011, Froment2018}. A key question is which spatial profiles of heating or energy input are capable of producing this imbalance, and whether they preferentially lead to prominence-like condensations, coronal-rain condensations, or both. Numerous studies have explored parameterised heating functions localised near the footpoints of coronal loops \citep{Antiochos1991, Antiochos1999, Muller2003, Muller2004, Johnston2019, Xia2011}, including the effects of asymmetry \citep{Mikic2013, Froment2018, Pelouze2022}. Others have investigated time-dependent, spatially random heating events, including impulsive heating and scenarios intended to mimic reconnection-driven nanoflares \citep{Parker1988, Karpen2008, Antolin2010, Jercic2023, Zhao2022}. The dynamics of these structures have been revealed through high-resolution modelling and multiwavelength observations. \cite{Jenkins2021} utilised extreme numerical resolutions (down to a 31 km grid cell size) to demonstrate that high resistivity allows prominence material to \textquotedblleft slip \textquotedblright\ across magnetic field lines, causing a slow descent of the entire body of the prominence. \cite{Oliver2014} modelled the kinematics of these falling blobs, finding they typically reach a constant terminal velocity (which is observed to be lower than free fall) when gravity is balanced by gas pressure gradients, while simultaneously exciting leaky sound waves during their descent. \cite{Moschou2015} captured the mass recycling process in 3D quadrupolar arcades, where evaporated plasma condenses into blobs that eventually fall back into the transition region. Statistically, \cite{Sahin2023} characterised observed quiescent coronal rain clumps as having widths (0.2-2 Mm) that remain constant with height, suggesting they are governed by fundamental heating and cooling scales. Extending this to active periods of the solar cycle, \cite{Sahin2024} observed flare-driven rain to be 4.4 times more abundant and significantly denser than quiescent rain, providing a direct link between solar activity, chromospheric evaporation and coronal rain. \cite{Wachira2026} observed the energetics of falling coronal rain, noting that coronal rain compresses the plasma ahead of it and generates hot rebound flows, reaching temperatures of 1.6-2.0 MK, upon impacting the lower transition region. These studies show that the temporal and spatial distribution of heating strongly influences the morphology and dynamics of the resulting condensations. However, the precise mechanisms that generate the required heating profiles, and whether these heating candidates, be it waves or reconnection, leave distinct thermodynamic signatures in coronal loops, remain unclear.

A wealth of observational and theoretical work has highlighted the ubiquity of Alfv\'en \citep{Alfven1942}, kink  and other magnetohydrodynamic (MHD) wave modes throughout the solar atmosphere \citep{Tomczyk2007, DePontieu2007, Morton2015, Morton2026, Ofman2015, Ofman2020, Ofman2023}. These waves can carry significant amounts of energy from the photosphere into the corona, where their dissipation may contribute to both coronal heating and solar wind driving \citep{Ofman2004, Ofman2010, Petrova2023, Lim2023, Lim2024}. Several numerical models have demonstrated that wave-driven heating can maintain coronal temperatures \citep{Ofman1998, Ballegooijen2011, Evans2012, McMurdo2026} and influence the structure of coronal loops \citep{Suzuki2005, Matsumoto2010}. In particular, \cite{Antolin2010} showed that Alfv\'en wave dissipation can suppress the formation of coronal rain due to the characteristic uniform heating they produce, which contrasts with footpoint-concentrated nanoflare heating models. However, the potential role of kink wave dissipation in triggering or shaping coronal condensations has not been as extensively studied. Additionally, the variable roles that Alfv\'en and kink wave heating have in perturbing the properties of the coronal plasma, thereby influencing the onset of thermal instability, remain incompletely understood. Each wave can dissipate its energy through several different mechanisms. In collisional plasmas, waves experience dissipative effects arising from the off-diagonal components of Braginskii’s pressure tensor \citep{Braginskii1965}, which account for viscous transport processes. Additionally, the magnetic diffusivity forces that appear in the generalised Ohm’s law \citep{Spitzer1962} work to dissipate both perpendicular and parallel currents. In addition to collisional dissipation mechanisms, Alfv\'en waves can also transfer their energy through nonlinear interactions between counter-propagating Alfv\'en waves. This drives Alfv\'en wave turbulence, enabling the cascade of energy to progressively smaller spatial scales where it can be efficiently dissipated \citep{Iroshnikov1964, Kraichnan1965, Howes2008}. \cite{Hossain1995} derived an energy decay model which, for non-zero cross-helicity, predicts a phenomenological dissipation term for homogeneous Alfv\'en wave turbulence, which was further refined by \cite{Matthaeus1999, Dmitruk2001}, and is now used extensively in Alfv\'en wave turbulence models, e.g., \cite{Holst2014, Downs2016, Reville2020}. This term replaces the nonlinear product of the counter-propagating Elsässer fields \citep{Elsasser1950} in the governing equations with an algebraic expression that is often assumed to be the Alfv\'en wave heating rate. 

Magnetised coronal plasma is highly structured, and transverse inhomogeneity alters how MHD waves propagate and dissipate. In the presence of inhomogeneity, Alfv\'en waves undergo phase mixing, which enhances dissipation via standard transport mechanisms such as shear viscosity and magnetic diffusion \citep{Heyvaerts1983}, and in a partially ionised plasma, ambipolar diffusion \citep{McMurdo2023, McMurdo2025}. Perturbations that propagate in an inhomogeneous plasma can also exhibit mixed properties when crossing Alfv\'en speed gradients \citep{Goossens2019}, leading to waves often being described as Alfv\'enic. \cite{Edwin1983} derived the dispersion relation for MHD waves in a cylindrical magnetic flux tube and identified, among others, the kink wave. While Alfv\'en waves are transverse (or torsional) oscillations of magnetic field lines, kink waves correspond to a collective transverse displacement of a magnetic flux tube. The restoring force remains predominantly magnetic tension, making the kink wave closely related to the Alfv\'en wave, but modified by the presence of transverse density structuring. In the thin tube limit, the kink wave can be interpreted as a surface Alfv\'en wave whose energy is concentrated near the boundary of the flux tube \citep{Goossens2009, Goossens2011}. Unlike Alfv\'en waves, kink waves that propagate along or against the magnetic field cannot be split into forward and backwards propagating Elsässer fields. \cite{Magyar2019} analytically demonstrated that perturbations transverse to the magnetic field, propagating through an inhomogeneous plasma, couple to compressible modes, causing each unidirectionally propagating perturbation to possess both Elsässer components. As is the case for counter-propagating Elsässer fields, co-propagating Elsässer fields also interact nonlinearly, as also shown by \cite{Magyar2017}, who produced simulations of an inhomogeneous plasma that exhibits a clear cascade where coherent density structures transition to a turbulent-like state. This effect, which has been described as a generalised form of phase mixing, was later termed uniturbulence by \cite{Magyar2019}. \cite{Doorsselaere2020} derived an energy cascade rate for both propagating and standing kink waves in terms of Elsässer fields, and showed their damping rate depends on the density contrast of the flux tube and the background plasma and is inversely proportional to the amplitude of the kink wave. \cite{Doorsselaere2024} later introduced the $Q$-variable, which allows for the decomposition of kink waves that propagate with or against the magnetic field, enabling \cite{Doorsselaere2025} to derive an evolution equation for kink waves that describes the energy dissipation and kink wave heating rate via uniturbulence. This allowed for transverse structuring to be incorporated within wave turbulence heating models, allowing for a multi-mode investigation to be performed. This system of equations derived by \cite{Doorsselaere2025} constitute the Uniturbulence and Alfv\'en Wave Solar Model (UAWSoM), which was implemented within the \texttt{MPI-AMRVAC} framework \citep{Keppens2012,Porth2014,Xia2018, Keppens2023} by \cite{McMurdo2026}, who found substantial differences in the ability of each wave mode to maintain a stable open-field coronal atmosphere. Namely, kink waves were able to maintain a stable open-field coronal atmosphere, in contrast to Alfv\'en waves, which rely purely on reflection driven heating. 


This investigation is structured as follows: in Section \ref{sec:Numerical Modelling}, we introduce the set of equations encompassed in UAWSoM and the magnetic field profile chosen for our coronal loop setup. In Section \ref{sec:Results} we present the results of our simulations for a range of parameterised and wave heating mechanisms. In Section \ref{sec:Discussion} we discuss our results and in Section \ref{sec:Summary and Conclusions} we present our conclusions and the implications of our results on future work. 

\section{Numerical methods}\label{sec:Numerical Modelling}

The $Q$-variables \citep{Doorsselaere2025} are defined as $\vec{Q} = \vec{V} \pm \alpha \vec{B}$, where $\vec{V}$ is the plasma velocity, $\vec{B}$ is the magnetic field, and $\alpha$ is a placeholder used to set the phase speed of the wave of interest. In the case of modelling kink waves, the quantity $\alpha B$ is taken to be the kink speed. This allowed \cite{Doorsselaere2025} to derive an evolution equation for kink wave energy evolution, analogous to those derived for Alfv\'en waves \citep{Holst2014}. This means we can now incorporate the uniturbulent dissipation and heating generated by kink waves in models of the solar corona without the typical harsh numerical constraints of having to resolve the length scales of turbulence. The system of equations that constitute UAWSoM is given by the standard MHD equations, modified by wave energy and pressure, combined with additional energy equations that govern the evolution of Alfv\'en and kink wave energy, and are given by

\begin{equation}
\frac{\partial \rho_0}{\partial t} + \nabla \cdot (\rho_0 \vec{V}) = 0,
\label{eq:continuity}
\end{equation}

\begin{equation}
\frac{\partial \vec{B}_0}{\partial t} - \nabla \times (\vec{V} \times \vec{B}_0) = 0,
\end{equation}

\begin{align}
\frac{\partial (\rho_0 \vec{V})}{\partial t}
+ \nabla \cdot \left( \rho_0 \vec{V} \vec{V}
- \frac{1}{\mu} \vec{B}_0 \vec{B}_0 \right)
+ \nonumber \\ \nabla \left( p + \frac{B_0^2}{2\mu}
+ P_\mathrm{A} + P_\mathrm{k} \right)
= -\rho_0 \vec{g}(\vec{y}).
\end{align}
Here, gravity is denoted by $\vec{g}$, and is given by 
\begin{equation}
    \vec{g}(y) = 274 \text{m}\,\text{s}^{-2}\frac{R_\sun^2}{\left(R_\sun + y\right)^2}\ \vec{1}_y,
\end{equation}
where $y$, is the vertical distance above the solar surface and $R_\sun$ is the solar radius taken to be $696.1$ Mm. The total energy equation is given by

\begin{align}
& \frac{\partial}{\partial t} \left(
\rho_0 \frac{V^2}{2}
+ \frac{p}{\gamma - 1}
+ \frac{B_0^2}{2\mu}
+ \sum W_\mathrm{A,k}^\pm
\right)
+ \nonumber \\ & \nabla \cdot \Bigg(
\left[
\rho_0 \frac{V^2}{2}
+ \frac{p}{\gamma - 1}
+ \frac{B_0^2}{2\mu}
\right] \vec{V}
- \vec{B}_0 \frac{\vec{V} \cdot \vec{B}_0}{\mu} - \kappa \nabla T
\Bigg)
+ \nonumber \\ & \nabla \cdot \left(
\vec{Z}_0^- W_\mathrm{A}^+
+ \vec{Z}_0^+ W_\mathrm{A}^-
+ \vec{Q}_0^- W_\mathrm{k}^+
+ \vec{Q}_0^+ W_\mathrm{k}^-
\right)
+ \nabla \cdot \left(
[P_\mathrm{A} + P_\mathrm{k}] \vec{V}
\right) \nonumber \\ &
= - \mathcal{L}
- \rho_0 \vec{V} \cdot \vec{g}(\vec{y})
+ \frac{\zeta - 1}{\zeta + 1}
P_\mathrm{k} \nabla \cdot \vec{V},
\label{eq:systemenergy}
\end{align}
where $\kappa = \kappa_0 T^{5/2}$, with $\kappa_0 = 8 \times 10^{-7}\ \mathrm{erg\ cm^{-1}\ s^{-1}\ K^{-7/2}}$. We neglect the effects of perpendicular thermal conduction, as, in the corona, it is typically several orders of magnitude lower than the parallel component due to the strong guiding influence of the magnetic field in the corona \citep{Braginskii1965}. The Alfv\'en and kink wave energy evolution equations are given by

\begin{equation}
\frac{\partial W_\mathrm{A}^\pm}{\partial t}
+ \nabla \cdot (\vec{Z}_0^\mp W_\mathrm{A}^\pm)
+ \frac{W_\mathrm{A}^\pm}{2} \nabla \cdot \vec{V}
= - \Gamma^\mp W_\mathrm{A}^\pm \mp \mathcal{R}_\mathrm{A},
\label{eq:alfven}
\end{equation}

\begin{equation}
\frac{\partial W_\mathrm{k}^\pm}{\partial t}
+ \nabla \cdot (\vec{Q}_0^\mp W_\mathrm{k}^\pm)
+ \frac{W_\mathrm{k}^\pm}{2} \nabla \cdot \vec{V}
= - \frac{1}{L_{\perp,\mathrm{VD}}}
\frac{1}{\sqrt{\rho_\mathrm{e}}}
(W_\mathrm{k}^\pm)^{3/2}\mp \mathcal{R}_\mathrm{k},
\label{eq:kink}
\end{equation}
where $\mu$ is the permeability constant, $p$ the hydrodynamic pressure. Derivations of the Alfv\'en and kink wave pressure contributions, given by $P_{\mathrm{A,k}}$, can be found in, e.g., \cite{Chandran2009, Doorsselaere2025, McMurdo2026}, and are given by

\begin{equation}
P_\mathrm{A}=\frac{W^+_\mathrm{A}+W^-_\mathrm{A}}{2} \quad \mbox{and} \quad
P_\mathrm{k}=\frac{W^+_\mathrm{k}+W^-_\mathrm{k}}{2\mu\rho_0\alpha^2}.
\label{eq:wave_pressure}
\end{equation}
The ratio of specific heats is set to $\gamma = 5/3$, appropriate for an ideal monatomic gas. The quantities $W_{\mathrm{A,k}}^\pm$ represent the Alfv\'en and kink wave energy propagating against and with the magnetic field, respectively. The vectors $\vec{Z}_0 = \vec{V}\pm\vec{v_\mathrm{A}}$ and $\vec{Q}_0=\vec{V}\pm\vec{v_\mathrm{k}}$ denote the equilibrium Elsässer and $Q$-variables associated with the Alfv\'en and kink speeds, respectively, with $\vec{B}_0$ corresponding to the equilibrium magnetic field. The term $\mathcal{L}$ accounts for the radiative losses, $\zeta = \rho_i/\rho_e$, the hypothetical density contrast, relating the ratio between the interior and exterior of the fine-scale coronal threads, and $\Gamma^\pm$ are the Alfv\'en-wave heating rates, which are introduced below. In addition, the perpendicular correlation length scales for Alfv\'en and kink waves are given by $L_{\perp,\mathrm{AW,VD}}$, where AW refers to Alfv\'en waves and VD refers to kink waves \citep[whose correlation length was derived by][]{Doorsselaere2025}. $\mathcal{R}_{\mathrm{A,k}}$ denotes the wave-energy reflection terms for each wave mode. In Equations (\ref{eq:continuity}-\ref{eq:wave_pressure}), the density $\rho_0$ is a weighted average of the external density, $\rho_\mathrm{e}$ and internal density, $\rho_\mathrm{i}$, of the assumed transverse structure, such as a plumelet or fine-scale coronal loop threads. The Alfv\'en wave heating terms are taken from \cite{Holst2014}, and are given by 

\begin{equation}\label{eq:AW_gamma}
    \Gamma^\mp = \frac{2}{L_{\perp,\mathrm{AW}}}\sqrt{\frac{W_\mathrm{A}^\mp}{\rho_0}},
\end{equation}
with $L_{\perp,\mathrm{AW}} = 1.5\times 10^9 B_{\mathrm{total}}^{-1/2}$, to remain consistent with \cite{Holst2014}. In the given expression, $B_{\mathrm{total}}$ is the total magnetic field, given in Gauss, is dependent spatially on $x$ and $y$, and the preceding constant has the appropriate units such that the total expression is given in centimeters. \cite{Doorsselaere2025} derived a correlation length for kink waves that is dependent on the physical structure of the plasma, and is shown to vary with the density contrast, filling factor, $f$, and the radius of the fine-scale coronal threads, $R$, and is given by
\begin{equation}\label{eq:LperpVD}
    L_{\perp,\mathrm{VD}}(R,\zeta,f) = \left(\frac{\sqrt{2}(\zeta - 1)}{2R\sqrt{5f\pi}} \frac{1-f^{5/2}}{(\zeta+1-f)^{3/2}}\right)^{-1}.
\end{equation}


We note that Equations (\ref{eq:alfven}-\ref{eq:kink}) do not allow for the two differing wave types to interact, thereby missing processes such as mode coupling \citep{Ruderman2013b, Shoda2018, Shoda2019, Goossens2019, Guo2019} and resonant absorption \citep{Ruderman2002, Goossens2002}. However, \cite{McMurdo2026} demonstrated that uniturbulence operates on shorter timescales than resonant absorption, for a wide range of parameters typical of Alfv\'en and kink waves observed in the solar corona, thereby leaving resonant absorption of second order importance when compared with the dissipation mechanisms considered. 

Kink waves require there to exist a density contrast between the interior and exterior of the fine-scale structures within the coronal loops \citep{Edwin1983}. In the present investigation, we take this density contrast to be 
\begin{equation}
    \zeta(y) = \zeta_0 \exp\!\left(-\frac{y}{5R_\odot}\right) + 1,
\end{equation}
with $\zeta_0 = 5$, resulting in density contrasts that fit within the range of observed values \citep{Aschwanden2003, Asensio2013}. The scale height for the density contrast is taken from previous investigations of UAWSoM \citep{Doorsselaere2025, McMurdo2026}, and have been retained in the present investigation. We assume that our density contrast varies solely with height, i.e., we do not consider the density contrast to vary along the loop, since this would require field line tracing at each time step and would rapidly increase the numerical resources required. The scale height of our density contrast is large compared to the dimensions of the box, meaning that it would not significantly affect the results. The calculation of the radiative losses is based on the implementation of \cite{Hermans2021}, and the radiative loss function is given by
\begin{equation}\label{eq:radiative_cooling}
    \mathcal{L} = \frac{\langle\rho_0^2\rangle}{(1+4A_{\text{He}})^2m_p^2}\Lambda(T).
\end{equation}
Here, $\langle\rho_0^2\rangle$ is the weighted average of the square of the internal and external densities by the filling factor (taken to be $f=0.1$) and density contrast of the loop. Additionally, $A_{\text{He}}$ represents the ratio of helium to hydrogen abundance, set to 0.1, while $m_p$ is the proton mass and the cooling curve, $\Lambda(T)$, was calculated using the data from \cite{Colgan2008}. We refer the reader to \cite{McMurdo2026} for further details regarding the implementation of radiative losses considered in UAWSoM.

\subsection{Reflection rates}\label{sec:Reflection}

Unlike kink waves, in order for Alfv\'en waves to undergo a turbulent cascade, we require a counter-propagating counterpart to exist co-spatially. This is evident in the evolution equation for Alfv\'en wave energy (Equation \ref{eq:alfven}), where the heating terms possess both Alfv\'en wave energies. Therefore, according to these equations, in an open magnetic field configuration, without reflection, Alfv\'en waves cannot undergo turbulent dissipation. Inwardly propagating Alfv\'en waves can, however, be generated through non-WKB reflection \cite{Heinemann1980}, which occurs when Alfv\'en waves propagate in a stratified environment. The necessary inhomogeneities are typically associated with field-aligned density gradients resulting from the stratification of the solar atmosphere and hence, Alfv\'en speed \citep{Velli1993}. Due to the averaging processes that occur when deriving the evolution equation for Alfv\'en wave energy, frequency-dependent information is lost, and our wave energy variable does not undergo reflection as a natural consequence of solving the system of equations in Section \ref{sec:Numerical Modelling}. Therefore, reflection is included as an additional sink-source term in the evolution equations for wave energy, such that all the energy removed from, e.g., an outwardly propagating wave is fed into its inwardly propagating counterpart. For Alfv\'en waves, we take this reflection rate from \cite{Holst2014}, and it is given by 

\begin{equation}
\mathcal{R}_\mathrm{A} = \mathcal{R}_{\mathrm{imb}}\sqrt{W_\mathrm{A}^\pm W_\mathrm{A}^\mp}
\begin{cases}
1 - 2 \sqrt{\frac{W_\mathrm{A}^{-}}{W_\mathrm{A}^{+}}},
& 4 W_\mathrm{A}^{-} \le W_\mathrm{A}^{+}, \\

0,
& \dfrac{1}{4} W_\mathrm{A}^{-} < W_\mathrm{A}^{+} < 4 W_\mathrm{A}^{-}, \\

2 \sqrt{\frac{W_\mathrm{A}^{+}}{W_\mathrm{A}^{-}}} - 1,
& 4 W_\mathrm{A}^{+} \le W_\mathrm{A}^{-},
\end{cases}
\label{eq:reflection1}
\end{equation}

\begin{equation}
\mathcal{R}_{\mathrm{imb}} =
\sqrt{
\left( \mathbf{b} \cdot [\nabla \times \mathbf{v}] \right)^2
+
\left[ (\mathbf{v}_A \cdot \nabla) \log V_A \right]^2
},
\label{eq:reflection2}
\end{equation}
where $\mathbf{b}, \mathbf{v}$ and $\mathbf{v}_A$ are the unit vectors associated with the magnetic field, velocity field and Alfv\'en speed. The imbalance-dependent reflection rate, $R_{\mathrm{imb}}$ is determined by gradients in the Alfv\'en speed and field-aligned vorticity, and the final reflection rate is limited by the local dissipation rate. Longitudinal structuring can also generate reflected kink-wave components \citep{Gao2024}, although particular density and flux-tube expansion profiles may permit non-reflective propagation \citep{Ruderman2013, Petrukhin2015, Ruderman2024}. However, a reflection rate that quantifies the transmission and reflection rates of kink wave energy due to, e.g., atmospheric stratification, is yet to be derived for use in wave-turbulence models, such as UAWSoM. Thus, for kink wave reflection, we take the kink analogue of Equations (\ref{eq:reflection1} - \ref{eq:reflection2}), meaning we replace the Alfv\'en speed with the kink speed and the Alfv\'en wave energy with kink wave energy in Equations (\ref{eq:reflection1}-\ref{eq:reflection2}). We propose that this implementation provides an equal footing for Alfv\'en and kink waves to redistribute their wave energy across the atmosphere without major inequality. We investigated the relative importance of the two terms in Equation (\ref{eq:reflection2}), and found that the dominant contributor was due to field-aligned gradients in each wave's propagation speed. The magnitude of these terms differs by approximately 2-3 orders of magnitude at later stages in the simulation, with larger discrepancies occurring at earlier stages in the simulation, where rotation is less developed. One should be aware that this is not a general statement that can be assumed for higher dimensional models, since the constant guide field considered in 2.5D MHD simulations generally limits the generation of rotational motions. We note that \cite{Pelouze2023} showed that transverse waves in the solar atmosphere experience a cutoff, such that waves with periods of 500 s or less may still reach the corona by tunnelling through the transition region with little to no attenuation. It should therefore be possible to estimate both the fraction of waves excited by photospheric motions that reach the solar corona and the frequency composition of that transmitted population, for example by frequency-binning the data presented in \citep{Lim2023,Lim2024}. Such an analysis could also provide a useful validation of parameterised kink wave reflection rates in future studies.

\begin{figure}[htp]
   \centering \includegraphics[width=.45\textwidth]{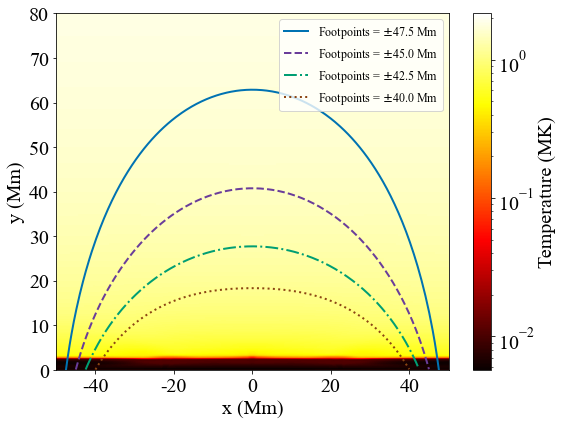}
    \caption{Representation of four magnetic field lines of our quadrupolar magnetic field topology at $t=1$ minute into our simulation, plotted over the temperature of the system given in MK. The footpoints of the field lines are located at $x = \pm 40, 42.5, 45$ and $47.5$ Mm.}\label{fig:Streamlines}
\end{figure}

As previously alluded to, the dimensions of our simulation are 2.5D, and include a chromosphere, transition region, and corona. The computational domain is a Cartesian box of $100 \times 80$ Mm with the $x$-axis in the range $[-50, 50]$ Mm and the $y$-axis (representing the vertical direction from the solar surface) initially in the range $[0, 80]$ Mm. It will be seen later that a larger box is required for one set of parameters. The temperature and density stratification are constructed in the same manner as \cite{Li2022} (see their Equations ~8 and 9). For the magnetic field, we adopt a quadrupole topology, similar to that used by \cite{Terradas2013, KeppensXia2014, Luna2016} and \cite{Jercic2024}. We follow the method of \cite{Zhang2019} by burying the magnetic null point by $y_0 = 4$ Mm, to improve numerical stability during the relaxation phase of the simulation. The background magnetic field is assumed time-independent, over which perturbations are allowed to evolve, and is given by

\begin{equation}
B_x = B_{\mathrm{base}} \cos(k_x x)e^{-k_x(y-y_0)} - B_{\mathrm{base}} \cos(3k_x x)e^{-3k_x(y-y_0)},
\end{equation}
\begin{equation}
B_y = -B_{\mathrm{base}} \sin(k_x x)e^{-k_x(y-y_0)} + B_{\mathrm{base}} \sin(3k_x x)e^{-3k_x(y-y_0)},
\end{equation}
\begin{equation}
B_z = B_{\mathrm{base}},
\end{equation}
where $B_{\mathrm{base}} = 10$ G, $k_x = \frac{\pi}{2L_0}$, and $L_0 = 50$ Mm, representing half the width of the domain. Given this description, throughout most of the domain the magnetic field strength is approximately $10$ G, while at the footpoints it reaches about $18$ G. A visual representation of the magnetic field is given in Figure \ref{fig:Streamlines}.

Coronal loops are, of course, not 2.5D, and the effects of the third coordinate are limited in their inclusion in this model through the use of an out-of-plane guide field. Coronal loops are also not perfectly symmetric, as considered in the present investigation, however, these considerations allowed for extensive testing with relatively inexpensive resources being required to evolve the simulations for sufficient time to develop either a prolonged steady state or develop condensations. As a proof-of-concept study, our aim here is to present the contrasting roles that the two waves can have in a setup where competing effects such as asymmetry are not considered. Further discussion on these points is given in \ref{sec:Summary and Conclusions}.

\subsection{Code details}

We solve the UAWSoM equations using the adaptive mesh refinement capabilities of \texttt{MPI-AMRVAC}. In the present work, three refinement levels were employed, which provided sufficient resolution for the structures of interest while deliberately avoiding unnecessary refinement of small transverse perturbations \citep[as was found in, e.g.,][]{Jercic2024} that could otherwise be interpreted as kink-like motions which might result in a double-counting of waves when combined with our UAWSoM wave energy formulation. Time integration was carried out with a three-stage SSPRK3 scheme, and the spatial fluxes were computed using an HLL approximate Riemann solver together with a second-order TVD slope limiter \cite{vanleer1974}. To improve the numerical robustness of our simulations, the magnetic field was treated using a field-splitting approach, in which the static background component is handled separately from the evolved perturbation. Because the simulations include both chromospheric and coronal plasma, together with a transition region, we also applied the transition region adaptive conduction (TRAC) method to better capture the steep thermal gradients. Control of numerical magnetic field divergence errors was achieved using a combined magnetic field divergence cleaning strategy available in \texttt{MPI-AMRVAC}. At the lateral boundaries, symmetric conditions were imposed, with antisymmetric conditions applied to the appropriate velocity and magnetic-field components. The lower boundary was treated as line-tied, with reflective velocity conditions, fixed thermodynamic variables consistent with hydrostatic equilibrium, and vanishing perturbations in the magnetic field. In addition, our wave energy injections were set constant in the lower boundary condition (more details in Section \ref{sec:ResultsWaves}). At the upper boundary, pressure and density were extrapolated according to the gravitational stratification, the velocity remained reflective. Lastly, the magnetic field was extrapolated with a second-order zero-gradient scheme. These choices were made to match the work carried out by \cite{Jercic2024}, to allow for a meaningful comparison between the differing employed heating mechanisms.

\section{Results}\label{sec:Results}

\subsection{Simulations without waves}\label{sec:ResultsNoWaves}

To establish whether Alfv\'en and kink wave heating produce thermodynamically distinct signatures in coronal loops, we first consider three baseline simulations that represent previously used heating prescriptions in coronal loop modelling.

\subsubsection{No externally applied heating}\label{sec:Noheating}

To isolate the role of heating in maintaining the simulated atmosphere, we first examine a simulation run in which no additional heating terms are included. In this case, both the Alfv\'en and kink wave energy terms are set to zero, and no background or localised heating is applied. The atmosphere evolves under the influence of the remaining physical processes, including thermal conduction, radiative losses, and the hydrodynamic response due to the imposed atmospheric stratification. The assumed density contrast necessary for kink waves to propagate is retained, so the radiative losses remain determined by the parameterised fine-scale structure of the plasma. As expected, the absence of compensating heating leads to a gradual cooling on a timescale typical of the solar corona \citep{DeMoortel2004, Aschwanden2008}.

\begin{figure*} 
    \centering
    \includegraphics[width=.99\textwidth]{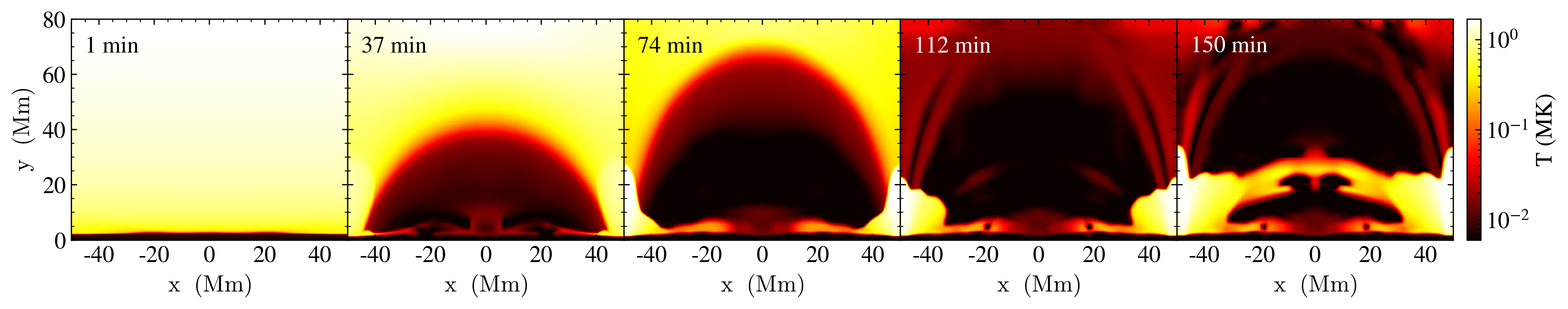}
    \caption{No heating: Spatial distribution of temperature from the 2.5D simulation within the computational domain at five representative times: $t=$ 1, 37, 74, 112, and 150 minutes.}
    \label{fig:No_heating}
\end{figure*}

Figure \ref{fig:No_heating} shows that the atmosphere cools preferentially in regions of weaker magnetic field strength. In the absence of heating, plasma flows and pressure gradients drive a redistribution of material toward lower heights, specifically toward the footpoints of the loop, leading to stronger adiabatic heating, helping to maintain comparatively higher temperatures in those regions. However, no prominences, coronal rain, or other strongly localised cool condensations form. This strongly supports the understanding that such structures require thermal non-equilibrium driven by an imbalance between heating and cooling, rather than cooling alone.

\subsubsection{Background heating}\label{sec:backgroundheating}

\begin{figure*} 
    \centering
    \includegraphics[width=.99\textwidth]{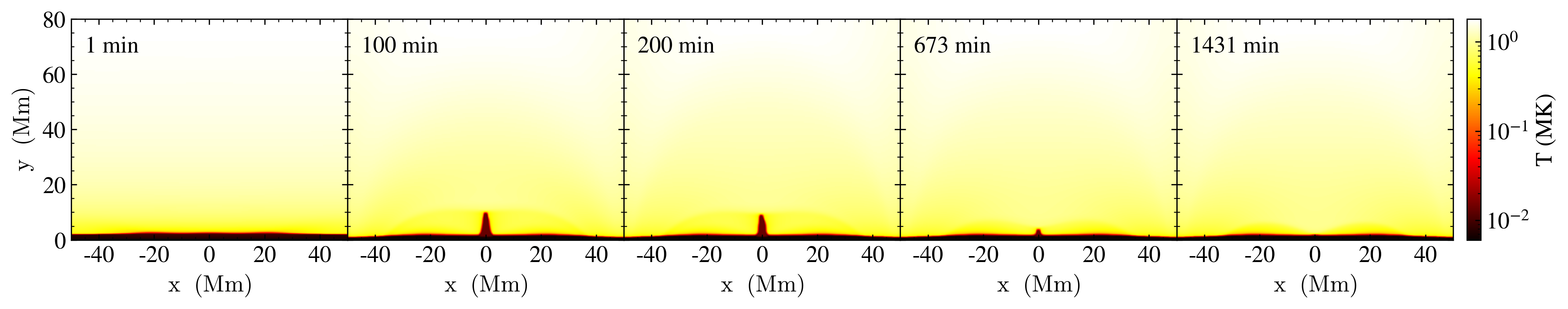}
    \caption{Background heating only: Spatial distribution of temperature from the 2.5D simulation within the computational domain at five representative times: $t=$ 1, 100, 200, 673 and 1431 minutes.}
    \label{fig:bQ_heating}
\end{figure*}

We now include an exponential background heating term, decreasing with height, given by
\[
H_0 \exp\left(-\frac{y}{\lambda_0}\right),
\]
with $H_0 = 10^{-4}\,\mathrm{erg\,cm^{-3}\,s^{-1}}$, and $\lambda_0 = 50\,\mathrm{Mm}$. These values were adopted from \cite{Jercic2024} and yield a stable atmosphere in which the heating balances the losses of the system. Figure \ref{fig:bQ_heating} shows the temporal evolution of the temperature of the atmosphere. The system initially advects the imposed initial conditions, after which it relaxes toward a quasi-steady state. The prominence-like structure forming in the centre of the domain originates from inward pressure gradients as the atmosphere relaxes. The appearance of this structure is aided by the particular magnetic field configuration and the fact that the null point is at the specific $x$-coordinate where the prominence-like structure forms.

\subsubsection{Background and localised heating}\label{sec:BLqheating}

\begin{figure*} 
    \centering
    \includegraphics[width=.99\textwidth]{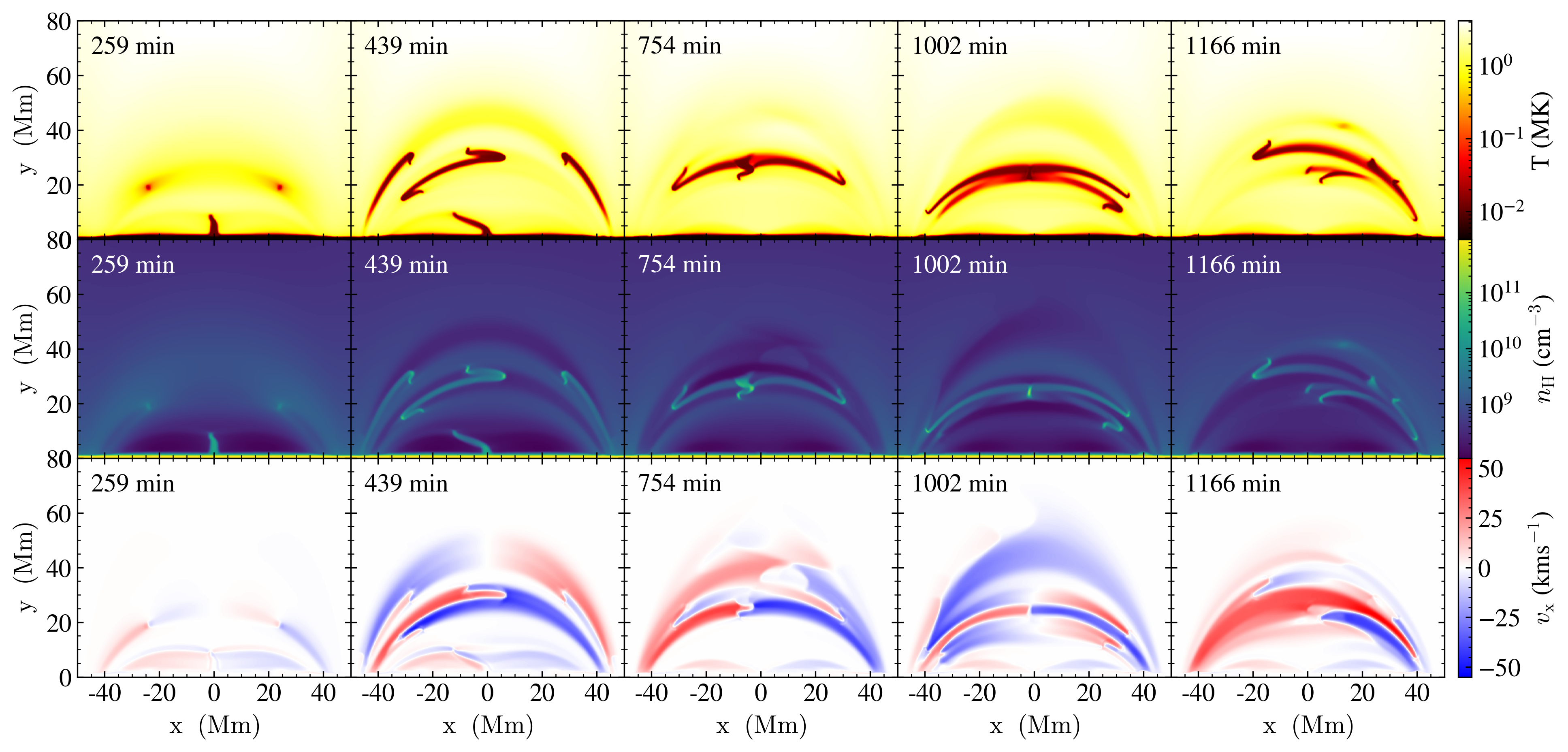}
    \caption{Background plus constant footpoint localised heating: Spatial distribution of temperature (top row), proton number densities (middle row) and the horizontal velocities (bottom row) at five representative times: $t=$ 259, 439, 754, 1002 and 1166 minutes.}
    \label{fig:bQ_lQ_heating}
\end{figure*}

In addition to the background heating term described above, we now include a steady, localised heating function concentrated near the footpoints of the coronal loop, following \cite{KeppensXia2014, Jercic2024}. This heating is introduced after the initial relaxation phase and is intended to represent an energy deposition representative of magnetic reconnection. Its functional form is given by $l_0$ multiplied by
\[
\begin{cases}
\exp\!\left[-\dfrac{(x-x_r)^2}{\sigma^2}\right]
+\exp\!\left[-\dfrac{(x-x_l)^2}{\sigma^2}\right], & y<y_h, \\[1.2em]
\exp\!\left[-\dfrac{y-y_h}{\lambda_h}\right]
\left(
\exp\!\left[-\dfrac{(x-x_r)^2}{\sigma^2}\right]
+\exp\!\left[-\dfrac{(x-x_l)^2}{\sigma^2}\right]
\right), & y\ge y_h,
\end{cases}
\]
where $l_0 = 2\times10^{-2}\,\mathrm{erg\,cm^{-3}\,s^{-1}}$, and $x_r=42$, $x_l=-42$, $y_h=4$, $\lambda_h=2.5$, and $\sigma^2=2$, are all given in Mm's. Figure \ref{fig:bQ_lQ_heating} shows the evolution of the temperature of the atmosphere and demonstrates the onset of cool condensations at 259 minutes (approximately 170 minutes after the localised heating term was introduced). Compared with our background-only heating case, a larger total energy input is injected per unit time step since the background heating is retained in addition to the localised footpoint contribution. Therefore, the dense structures form, not due to a lack of heating, but as a consequence of increased heating, suggesting that the severity of stratification is key when determining the stability of the plasma, in agreement with \cite{Froment2018}, who found that the more localised the heating is around the footpoints, the more likely the loop is to undergo thermal non-equilibrium (TNE) cycles. Notably, in our simulations, no stable prominence forms, as was the case in \cite{Jercic2024}, which we attribute to the increased radiative losses due to the assumed fine-scale structuring of thin dense structures within the coronal loop, captured by the $\zeta$ function, leading to a faster transition to instability. This led to a stronger cooling rate and shows that thread-like condensations can form in the steady localised heating case also, not just in the localised, stochastic heating case, in which the localised heating rate was spatially varying with time. 

The condensations in our simulation are at least two orders of magnitude denser than the corona, with values of the order of a few $10^{11}$ cm$^{-3}$. Due to this, these condensations experience strong gravitational forces resulting in them falling towards the chromosphere. Our simulations exhibit horizontal velocities of the order of $50$ kms$^{-1}$, with maximum total velocities reaching $>75$ kms$^{-1}$. This is roughly 1.5-2.5 times larger than observations of coronal rain \citep{Liu2012, Vashalomidze2022}. It should be noted that the range of values observed by, e.g., \cite{Antolin2023}, is rather large ($10 - 150$ kms$^{-1}$), and there likely exist differing regimes in which the values observed in our simulations are realistic of the solar corona. Perhaps the additional contributions from waves will slow these condensations down, and explain the situations in which we observe slower coronal rain. 

\subsection{Simulations with wave energy injections}\label{sec:ResultsWaves}

Having established the behaviour of the atmosphere in the absence of waves and under purely parameterised heating, we now examine the effects of Alfv\'en and kink wave heating on the thermodynamics of a simulated coronal loop. In the wave-driven cases, we adopt a single wave energy injection value of $1.59\,\mathrm{erg\,cm^{-3}}$, corresponding to chromospheric velocity perturbations of $3\,\mathrm{km\,s^{-1}}$. This value also results in a wave energy content reaching the corona that is consistent with values inferred from observations of kink waves \citep{Weberg2020}, providing further realism to the adopted dissipation lengths in UAWSoM. The injected wave energy is made dependent on the sign of the magnetic field, such that only wave energy that propagates from the surface into the domain is considered. This dependence is represented below by

\begin{equation}
\begin{cases}
W_\mathrm{A, k}^- = 1.59 \ \mathrm{erg\ cm^{-3}}, \qquad
W_\mathrm{A, k}^+ = \mathrm{Open},
& B_y > 0, \\[6pt]
W_\mathrm{A, k}^+ = 1.59 \ \mathrm{erg\ cm^{-3}}, \qquad
W_\mathrm{A, k}^- = \mathrm{Open},
& B_y < 0,
\end{cases}
\label{eq:wave_energy_bc}
\end{equation}
where \textquotedblleft Open\textquotedblright\ means that a zero-gradient boundary condition is used to allow out-flowing wave energy to leave the domain. In the following sections, we present the results when Alfv\'en or kink wave energy is injected into our domain, including simulations where a given correlation length is used and when this correlation length is reduced by a factor of ten, thereby reducing the scale height of the dissipation of each wave, to investigate a situation when there is more localised wave heating. For clarity, we present below the forms of the kink and Alfv\'en wave heating rates, given by $Q_{\mathrm{k}}$ and $Q_{\mathrm{A}}$, respectively.

\begin{equation}
    Q_{\mathrm{k}}^\pm = \frac{1}{L_{\perp,\mathrm{VD}}}\frac{1}{\sqrt{\rho_\mathrm{e}}}(W_\mathrm{k}^\pm)^{3/2}, \\
    Q_{\mathrm{A}}^\pm = \frac{2}{L_{\perp,\mathrm{AW}}}\sqrt{\frac{W_\mathrm{A}^\mp}{\rho_0}}W_\mathrm{A}^\pm.
\end{equation}
The total heating rate is given by the sum of the contributions from the counter-propagating (+ and -) wave components.

\subsubsection{Kink wave driven case} \label{sec:Kinkheating}

The radius of the fine-scale structures on which kink waves propagate is assumed to be proportional to the magnetic field as 
\begin{equation}
    R(x,y) = R_0 \sqrt{\frac{B_{\mathrm{base}}}{B_{\mathrm{total}}(x,y)}},
\end{equation}
where we take $R_0$ to be 1 Mm. Due to the transverse variation in the surface magnetic field, the base radius actually varies between ~0.75 and 1 Mm, with smaller values relating to regions where the magnetic field is stronger than 10 G.

For the kink-wave-driven case, we include a background heating term during the initial relaxation phase of the simulation to prevent the formation of condensations that occur due to the advection of the initial conditions. This background heating term is gradually reduced every fifteen minutes of simulated time before being switched off completely after one hour, leaving the kink wave heating as the sole energy input. We found that if the background heating term is not included during the initial phase, the simulation takes a considerable amount of time and numerical resources to relax to a near-identical state as that when a background heating term is applied. In the absence of this term, cool structures form, merge and eventually fall into the chromosphere, leaving a stable atmosphere. 

\begin{figure*} 
    \centering
    \includegraphics[width=.99\textwidth]{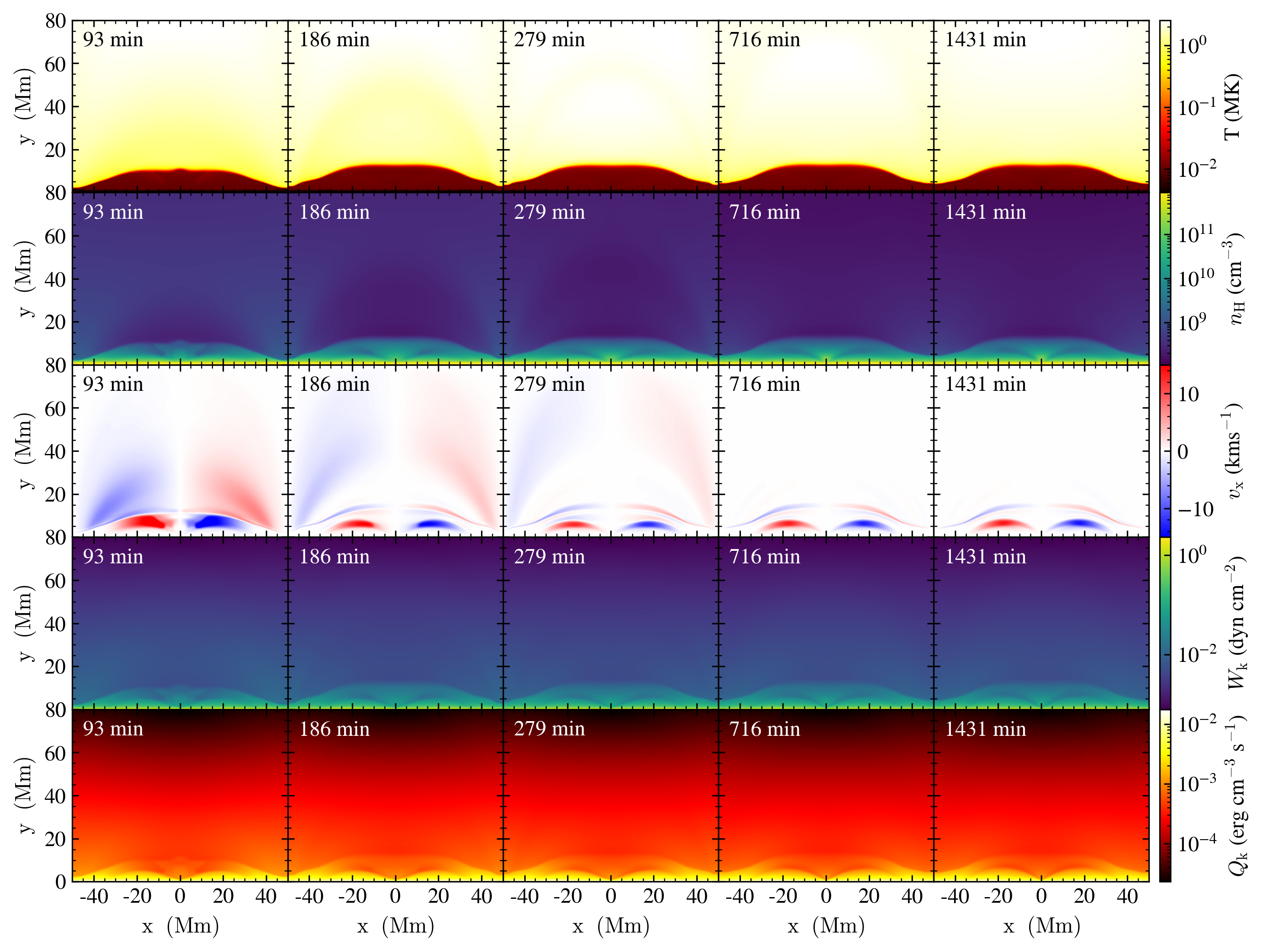}
    \caption{Kink wave energy injection: Spatial distribution of temperature (top row), proton number densities (second row), horizontal velocities (third row), total kink wave energy (fourth row) and the kink wave heating rate (bottom row) at five representative times: $t=$ 93, 186, 279, 716, and 1431 minutes.}
    \label{fig:Khom_Lperp1}
\end{figure*}
In Figure \ref{fig:Khom_Lperp1}, we present the full simulation domain at five different snapshots in time for the temperature (top row), proton number density (second row), horizontal velocity (third row), total kink wave energy (fourth row), and the kink wave heating rate (bottom row). The kink wave heating has a large enough scale height that it acts like a global heating rate (similar to the background-only heating case), resulting in a very stable atmosphere. The chromosphere is raised considerably, and this is attributed to the additional pressure and large kink wave heating occurring in the lower layers of the solar atmosphere. We also note that the horizontal velocity of the coronal plasma reduces in time, leading to a near-static plasma, suggesting the coronal portion has reached equilibrium. Our conclusion regarding this simulation is that the relatively large radius assumed for the fine-scale structures results in weakly stratified heating and consequently a very stable loop. 

\subsubsection{Alfv\'en wave driven case}\label{sec:AWheating}

\begin{figure*} 
    \centering
    \includegraphics[width=.99\textwidth]{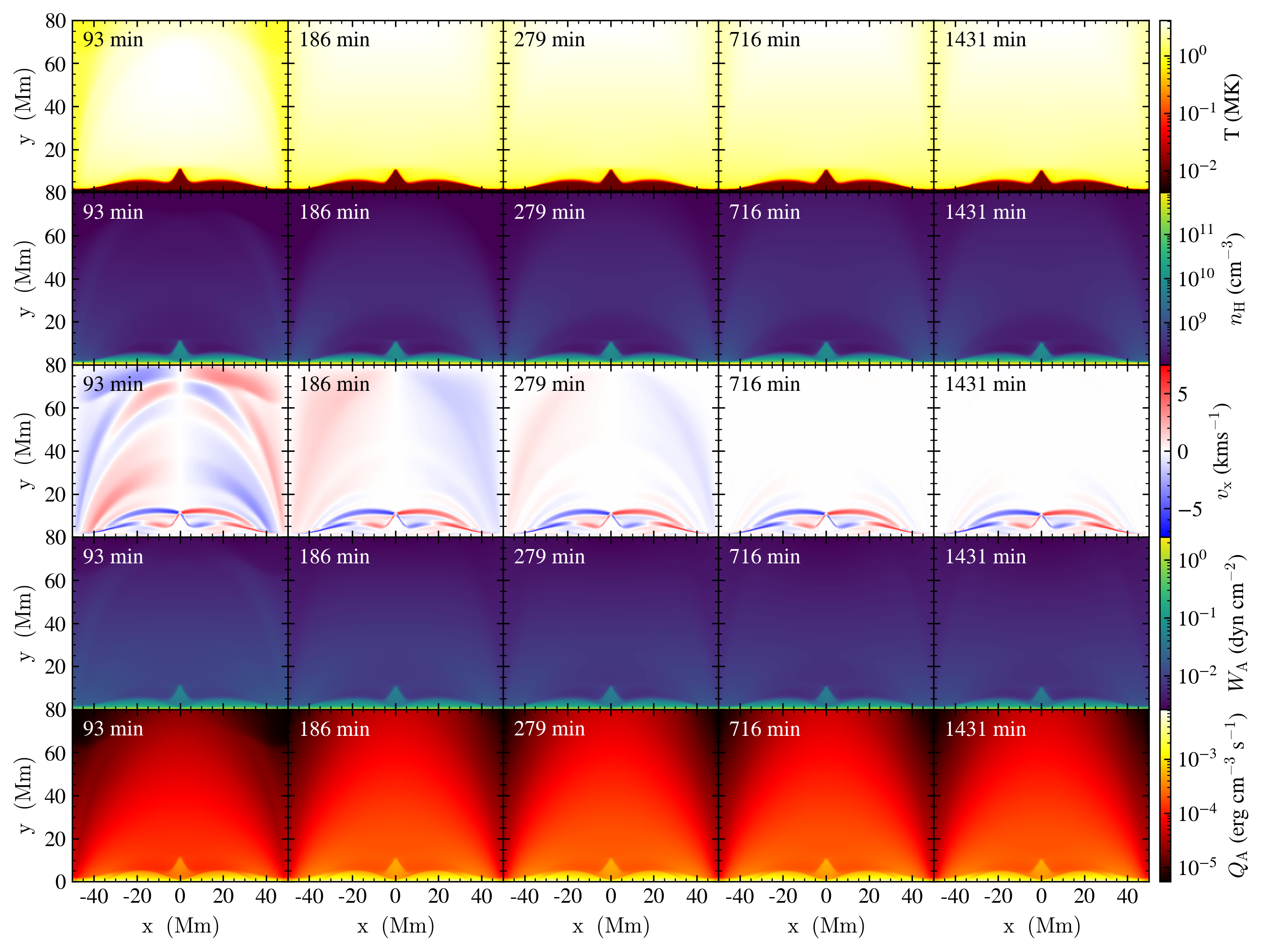}
    \caption{Alfv\'en wave energy injection: Spatial distribution of temperature (top row), proton number densities (second row), horizontal velocities (third row), the total Alfv\'en wave energy (fourth row) and the Alfv\'en wave heating rate (bottom row) at five representative times: $t=$ 93, 186, 279, 716, and 1431 minutes.}
    \label{fig:AWhom_Lperp1}
\end{figure*}

In the Alfv\'en wave driven case, we find that an additional background heating term is not required to maintain a stable atmosphere during the initial relaxation phase, and incorporating the same heating term makes no difference to the final solution. The dissipation of Alfv\'en wave energy provides a sufficiently stable energy input to offset the radiative losses and sustain the corona without producing condensations. As was the case in the kink wave heating simulation, the atmosphere undergoes an initial adjustment from the imposed initial conditions and then relaxes toward a quasi-steady state. This is clear from the third row of Figure \ref{fig:AWhom_Lperp1}, whereby the horizontal velocities are very small.

Figure \ref{fig:AWhom_Lperp1} shows that the Alfv\'en wave heating is also weakly stratified within the corona and acts similarly to the background-only (non-local) heating case presented in Figure \ref{fig:bQ_heating}. A small prominence/spicule-like structure is formed in the centre of the domain, although now it persists throughout the entire simulation. This can be attributed to the long-lasting horizontal velocities low in the domain, generated by additional pressure contributions from Alfv\'en waves, constantly driving mass towards the centre of the domain. Towards the end of the simulation, the heating profile shows a structure that traces out the global profile of the closed field lines, as shown in the bottom panel of Figure \ref{fig:AWhom_Lperp1}. Here, the heating is generated from Alfv\'en waves originating from uncorrelated sources, whereas in the upper left and right corners, the heating is generated through reflection alone, since these magnetic fields do not connect back within the domain. 

In general, the correlation lengths for Alfv\'en waves used here are large enough compared with the box dimensions and the lengths of the simulated coronal loops to maintain stability. The Alfv\'en and kink wave heating in these two cases is not sufficiently stratified to generate thermal instability, resulting in long-lasting stability of the atmosphere. It is therefore a natural question to ask whether reducing the scale height of the wave dissipation, through reducing each wave's correlation length, can trigger the formation of cool dense plasma.  

\subsection{Varying the correlation lengths for each wave}

Discrepancies between observations, theory and simulations suggest the correlation length of Alfv\'en and kink waves is not well constrained \citep{Ballegooijen2011}, with observed values differing by as much as two orders of magnitude \citep{Sharma2023}. We now present our findings when we reduce the correlation lengths of each wave (Equation \ref{eq:LperpVD} for kink waves and the expression following Equation \ref{eq:AW_gamma} for Alfv\'en waves) in our simulation by a factor of ten. We retain the standard functional forms for each correlation length and vary them by the same order of magnitude, rather than enforcing identical correlation lengths for each mode. This choice reflects the fact that each correlation length depends differently, or not at all, on various properties of the plasma. These distinctions are intrinsic to the wave mode under consideration, and it is therefore inappropriate to constrain all of these variables simultaneously. Variation in correlation length feeds into the dissipation terms in the wave energy evolution equations, which ultimately sets the heating in the atmosphere. 

\subsubsection{Alfv\'en wave driven case: reduced correlation length}

\begin{figure*} 
    \centering
    \includegraphics[width=.99\textwidth]{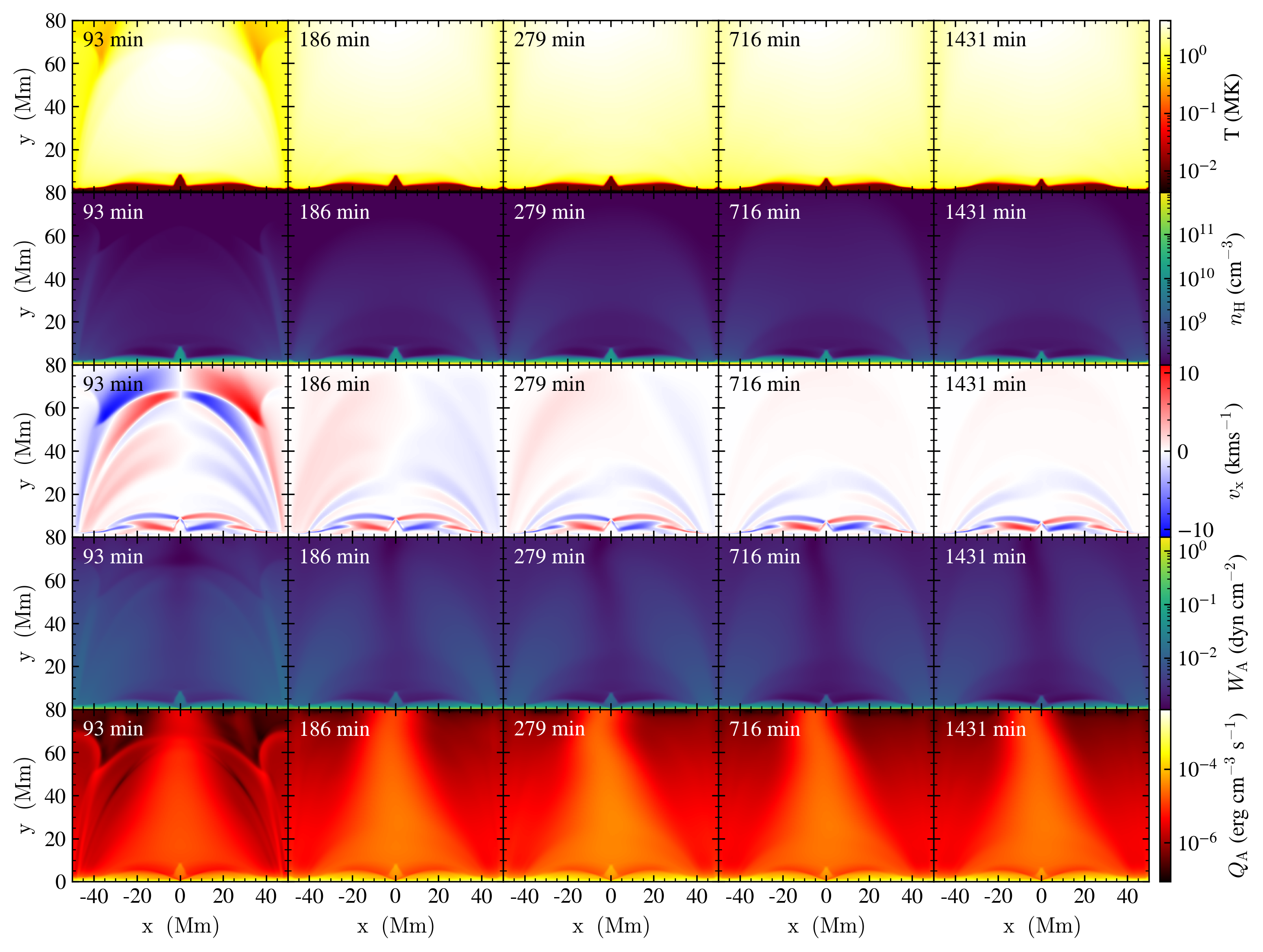}
    \caption{Alfv\'en wave energy injection with $L_{\perp,\mathrm{AW}}$ reduced by a factor of ten: Spatial distribution of temperature (top row), proton number densities (second row), horizontal velocities (third row), the total Alfv\'en wave energy (fourth row) and the Alfv\'en wave heating rate (bottom row) at five representative times: $t=$ 93, 186, 279, 716, and 1431 minutes.}
    \label{fig:AW_Lperp0p1}
\end{figure*}

We consider the same energy injection as previously, only now, the correlation length has been made ten times shorter. The top two panels of Figure \ref{fig:AW_Lperp0p1} show similar behaviour to the previous case for Alfv\'en waves. The distribution of plasma differs slightly, in that now the chromosphere is not as raised as was the case when the original correlation length was considered. The wave heating profile (bottom panel) differs quite significantly, however, and this can be attributed to the more highly stratified nature of Alfv\'en wave energy dissipation. The Alfv\'en wave heating is much more noticeably concentrated in the centre of the domain at the apex of each magnetic field line, with the overall profile resembling a triangle. This is to be expected as the Alfv\'en wave heating rate has been increased by reducing the correlation length and that balanced/two-footpoint wave driving tends to maximise the Alfv\'en wave turbulent cascade \citep{Downs2016}. 

\begin{figure} 
    \centering
    \includegraphics[width=.45\textwidth]{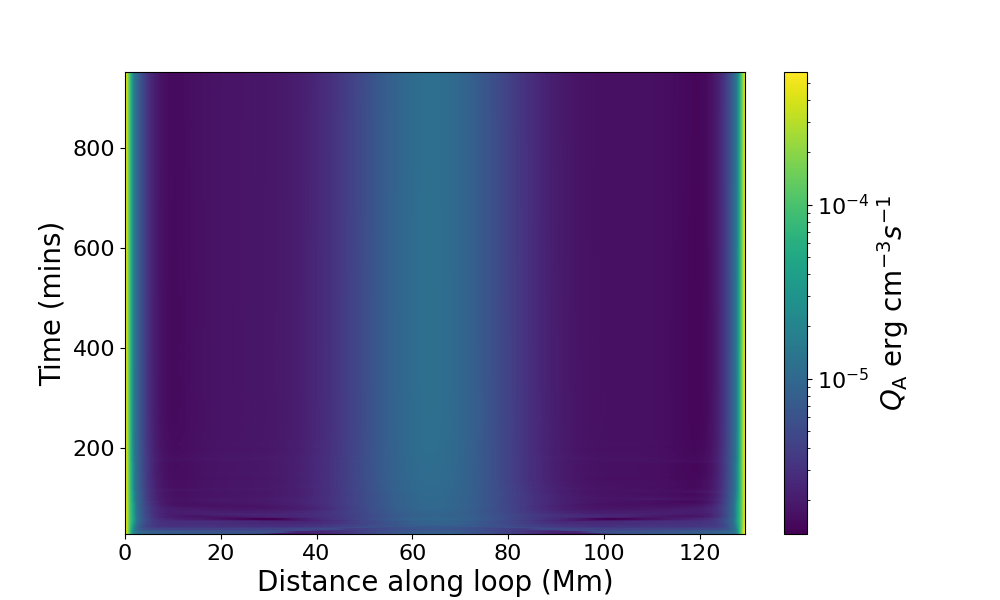}
    \caption{The time evolution of the Alfv\'en wave heating rate along a loop with footpoints at $x = \pm 45$ Mm, with $L_{\perp,\mathrm{AW}}$ reduced by a factor of ten.}
    \label{fig:AW_Lperp0p1_fieldline_m4p5}
\end{figure}

This triangular profile was seen to a lesser extent in the previous Alfv\'en wave driven simulation, displayed in Figure \ref{fig:AWhom_Lperp1}, and is now enhanced by the reduction in correlation length. Notably, the atmosphere remains very stable, suggesting Alfv\'en wave heating still acts as a very stable heating mechanism. Only at lower altitudes do we see larger horizontal velocities driving mass towards the centre of the domain. The velocities are roughly twice as large compared with the case where the original value for the Alfv\'en wave correlation length is considered, and this is attributed to the more localised dissipation of Alfv\'en wave energy occurring due to the reduced correlation length. Stability remains in the Alfv\'en-wave-driven case because there exists strong loop-top heating, which is known to result in very stable heating, as it counterbalances the typical evaporation and localised (close to) loop-top cooling and condensation formation typical of coronal rain formation \citep{Antiochos1999}. Figure \ref{fig:AW_Lperp0p1_fieldline_m4p5} shows a sampling of the wave energy and Alfv\'en wave heating rate along a single loop with footpoints at $x = \pm 45$ Mm. The 2D profile of this loop can be seen in the context of the whole domain in Figure \ref{fig:Streamlines}. Figure \ref{fig:AW_Lperp0p1_fieldline_m4p5} shows a clear enhancement in Alfv\'en wave heating at the loop top. This can be thought of as a collision site between the two Alfv\'en wave energy variables driven from opposite loop foot points. It is in this region where the most pronounced Alfv\'en wave dissipation and hence, the strongest heating occurs, resulting in a strong balancing force that ensures thermal stability of the plasma. Other loops in the domain exhibit similar behaviour, although the location of the enhanced loop top wave heating becomes more localised as the height of the loop apex increases, or equivalently, the loop footpoints become further apart.

\subsubsection{Kink wave driven case: reduced correlation length}

Variations in the kink wave correlation length can be motivated on a more physical basis than for Alfv\'en waves. The definition of $L_{\perp,\mathrm{VD}}$ clearly shows that the radius of the fine-scale structures plays a key role in determining the correlation length. Investigations by \cite{Morton2023} suggest that the corona is structured down to the smallest observable scales, implying that the emitting plasma may occupy significantly more than 10\% of the observed volume (i.e., a filling factor larger than 0.1), and, therefore, the radius of the fine-scale structures may be smaller than the finest resolved scale. \cite{Tajfirouze2026} estimated the coherence length of Alfv\'enic fluctuations from observations to be on the order of 300-500 km. We emphasize that this does not correspond to the outer scale that governs nonlinear interactions and energy transfer; rather, it may be interpreted as the full width (or twice the radius) of coherently oscillating flux tubes. For our chosen parameters, this implies a reduction of the originally assumed correlation length by a factor between four and seven. Furthermore, observations by \cite{Williams2020b, Williams2020} have resolved strand widths of the order of 200 km, using the High-Resolution Coronal Imager (Hi-C). Clearly, the characteristic radius of our fine-scale coronal threads need not be of the order 1 Mm, and may in fact be significantly smaller. With this in mind, we now present the results obtained from a simulation in which the correlation length for kink waves has also been reduced by a factor of ten from its initial value, thereby matching the smallest scales observed by \cite{Williams2020}. For our chosen values of the filling factor and the density contrast, the relationship between the correlation length for kink waves, as a function of radius, is well approximated by $L_{\perp,\mathrm{VD}} = 6.5 R$ (given in Mm), due to the comparatively large scale height of our density contrast with the loop length. Clearly, the radius of the fine-scale coronal strands in our coronal loops has a strong impact on the correlation length and hence kink wave heating rate. 


When performing simulations with a reduced correlation length, we found condensations formed (at approximately 40 Mm above the surface), and began to interfere with the upper boundary of the computational domain, indicating that a larger domain would be necessary to capture their further evolution. Therefore, we increased the vertical dimensions of our box by a factor of two. We retain the initial background heating function as used previously to allow the system to relax without immediately triggering condensations, along with all other previously given modelling specifics, including the same grid cell size, achieved by increasing the number of cells in our domain.

\begin{figure*} 
    \centering
    \includegraphics[width=.9\textwidth]{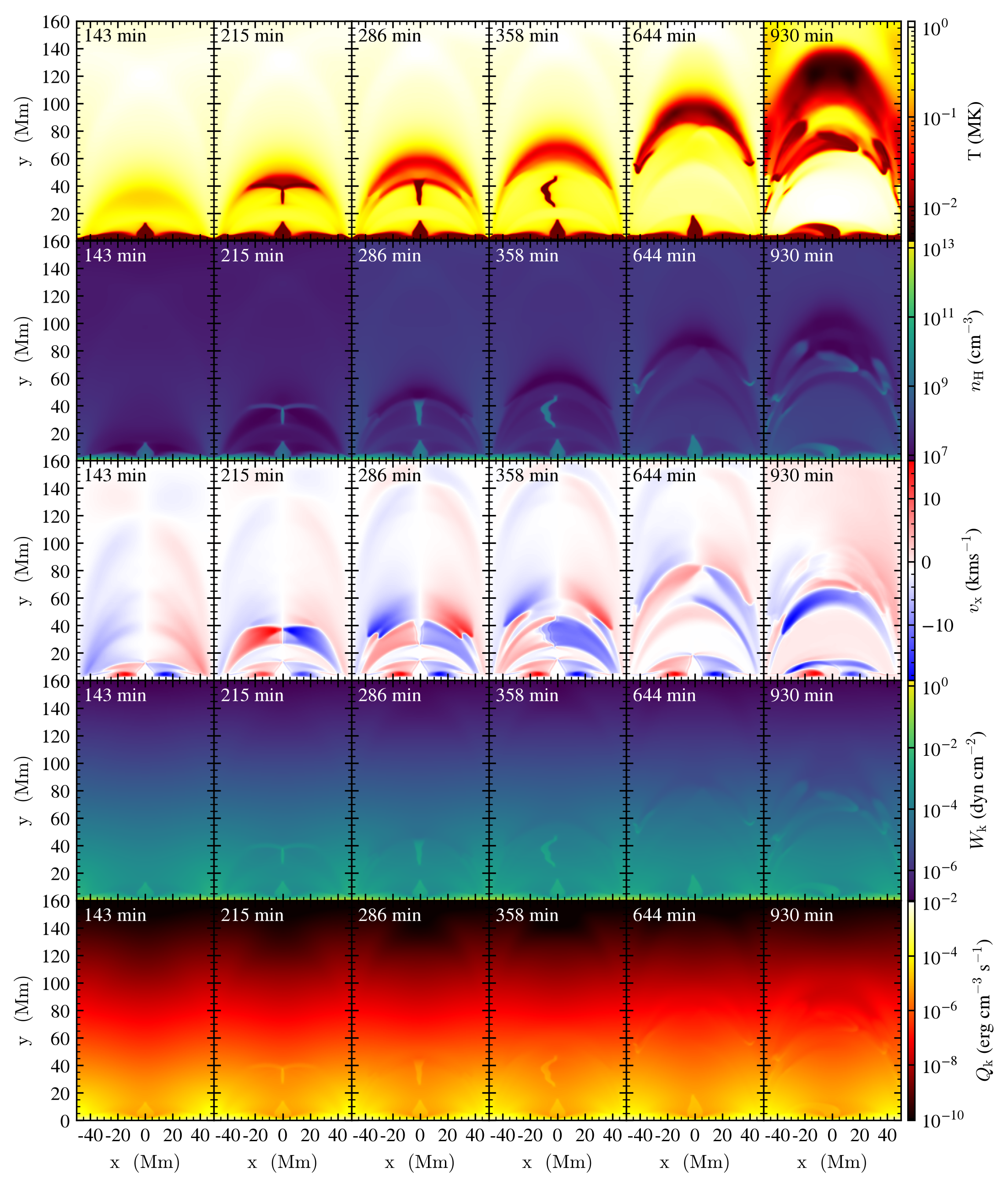}
    \caption{Kink wave energy injection with $L_{\perp,\mathrm{VD}}$ reduced by a factor of ten: Spatial distribution of temperature (top row), proton number densities (second row), horizontal velocities (third row), the total kink wave energy (fourth row) and the kink wave heating rate (bottom row) at six representative times: $t=$ 143, 215, 286, 358, 644 and 930 minutes.}
    \label{fig:dblK0p1_heating_Te}
\end{figure*}


As shown in Figure \ref{fig:dblK0p1_heating_Te}, at approximately 215 minutes, a cool, dense structure appears at the apex of the magnetic field lines (40 Mm in altitude). As the simulation progresses, it is clear that not all cool plasma is dense, as was the case in the simulation in Section \ref{sec:BLqheating}. We attribute the formation of cool dense structures to the profile of the kink wave heating rate shown in the bottom row of Figure \ref{fig:dblK0p1_heating_Te}. The colour scale has been fixed such that it does not represent the full range of values in the domain ($1.1 \times 10^{-2}$ - $5 \times 10^{-11}$ erg cm$^{-3}$s$^{-1}$), but rather to aid the following explanation. The dominant heating in our domain is occurring at the loop footpoints, which contrasts with the heating from Alfv\'en waves, which occurs at the loop apexes. With kink waves, we have reproduced a heating rate that reflects much more closely the localised footpoint heating that we imposed as a parameterised function. By reducing the radii of the fine-scale plasma structures within the coronal loop to be of the order of 100 km, we observe a stratified heating rate that causes condensations to form along the coronal loop. The decrease in temperature observed at later times is not accompanied immediately by a comparable increase in density. Previous numerical studies of prominence and coronal-rain formation have demonstrated that the initial nonlinear cooling phase can evolve approximately isochorically, with the temperature decreasing considerably faster than the density responds \citep{Xia2011, Moschou2015, Jenkins2021, Brughmans2022}. The cool, comparatively tenuous region in Figure \ref{fig:dblK0p1_heating_Te} may therefore represent an early stage of runaway cooling instead of a developed/developing condensation. A substantial density enhancement may develop subsequently, when pressure gradients drive mass toward the cooling region.

Our simulated condensations do not reach the same density as those in the locally parameterised heating case, owing to the additional contribution from kink wave heating having a restabilising contribution. Despite there being strong reflection of kink wave energy at the edges of the condensations, due to large gradients in wave speed, kink wave energy still accumulates around the condensations because the high density reduces the propagation speed of the wave, leading to a substantial kink wave heating rate. This is seen even more clearly below in Figure \ref{fig:field_line_plots} (bottom panel), in which the heating remains enhanced around the condensations. We omit the first few time steps of the kink wave heating data to remove the initial advection period of wave energy, such that the contrast allows for a clearer analysis. The accumulation of wave energy around the condensations is a natural consequence of wave energy transport and is not present in our local parameterised footpoint heating case, since the parameterised heating rates remain fixed in position, unlike the propagating wave energy. The maximum horizontal velocities shown in the third row of Figure \ref{fig:dblK0p1_heating_Te} are approximately $15$ kms$^{-1}$, and the magnitude of the total velocity of the falling condensations fall in the range of $15-30$ kms$^{-1}$, considerably smaller than those of the local parameterised heating simulation, showcasing the counteracting force of kink waves on balancing the cooling processes in the solar atmosphere to a greater extent. These values fall within the lower end of the observationally inferred range, suggesting that both localised heating mechanisms, potentially associated with magnetic reconnection, and kink wave heating may operate simultaneously. Alternatively, the results may indicate that a range of correlation lengths exists, in which case the observations could be explained entirely by kink wave heating if the fine-scale structures reach values smaller than 100 km. 

The various field lines shown in Figure \ref{fig:field_line_plots} depict very different density evolution because kink wave heating and radiative cooling are intrinsically non‑uniform throughout the domain. Small cross‑field variations in the kink wave dissipation and energy flux alter the density of the solar corona to varying degrees, leading to amplified effects of radiative losses. A field line that receives slightly different heating can become thermally unstable and form condensations, while neighbouring field lines with marginally shorter loops exhibit very different behaviour. Therefore, the evolution of the density along field lines with footpoints located at $x=\pm 40, 42.5, 45$ and $47.5$ Mm, shown in Figure \ref{fig:field_line_plots}, reflects the sensitivity of thermal stability to small spatial variations in kink wave energy deposition. This variation in behaviour is impossible to reproduce in traditional 1D loop simulations of coronal rain, which can only impose a specific heating rate along a single field line. Our 2.5D MHD model shows that even small cross‑field differences in heating lead to dramatically different thermal evolution and stability.

\begin{figure*}[htp]
    \centering

    \begin{minipage}[t]{0.2495\textwidth}
        \centering
        \includegraphics[width=\textwidth]{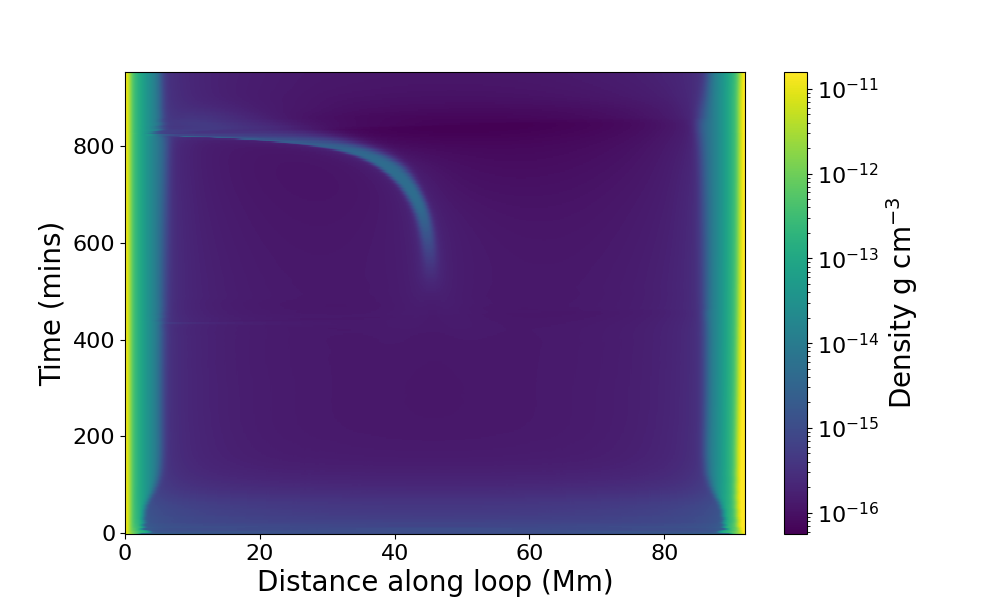}
    \end{minipage}\hfill
    \begin{minipage}[t]{0.2495\textwidth}
        \centering
        \includegraphics[width=\textwidth]{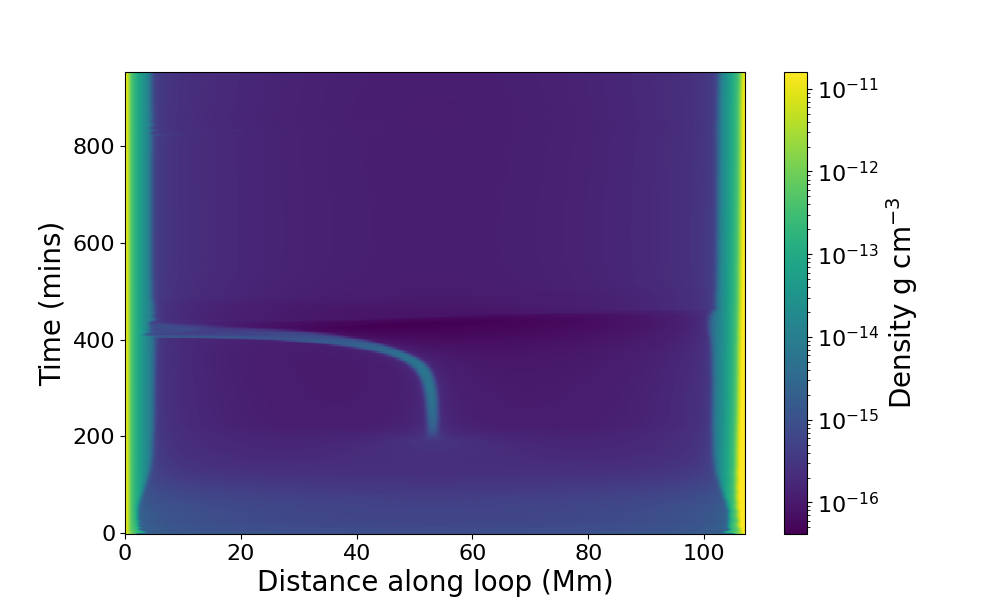}
    \end{minipage}\hfill
    \begin{minipage}[t]{0.2495\textwidth}
        \centering
        \includegraphics[width=\textwidth]{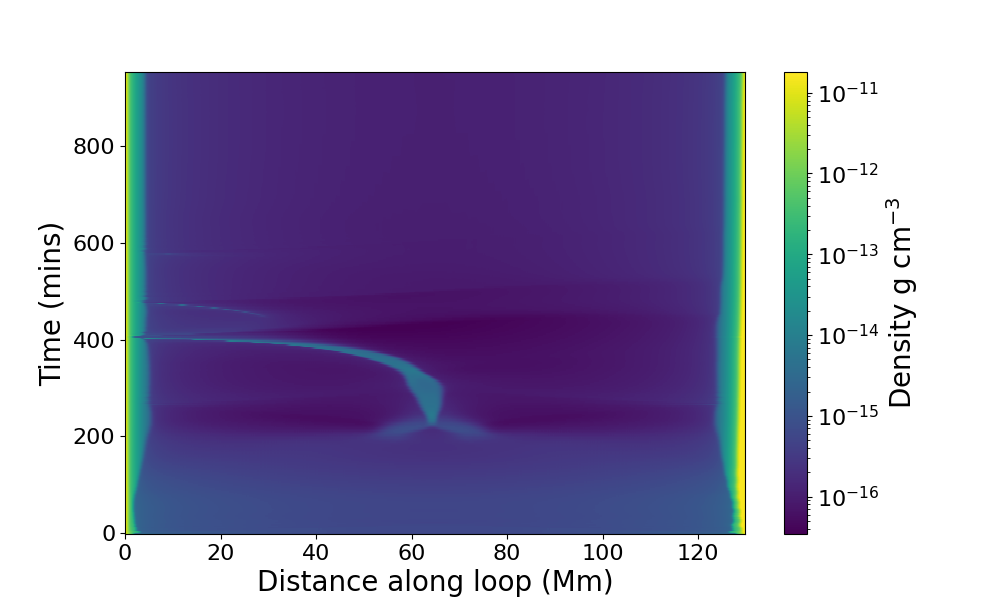}
    \end{minipage}\hfill
    \begin{minipage}[t]{0.2495\textwidth}
        \centering
        \includegraphics[width=\textwidth]{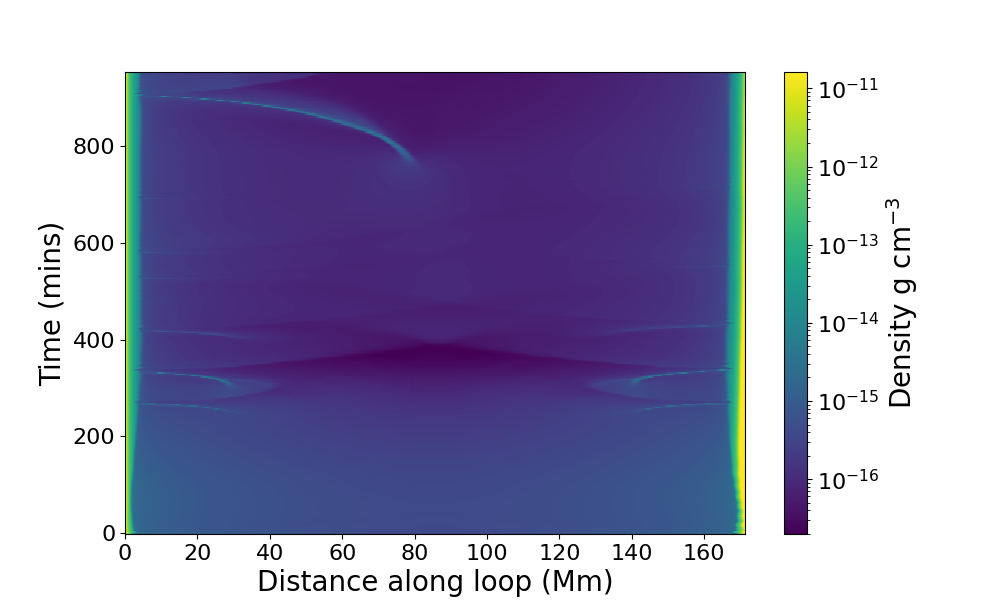}
    \end{minipage}

    \vspace{0.5em}

    \begin{minipage}[t]{0.2495\textwidth}
        \centering
        \includegraphics[width=\textwidth]{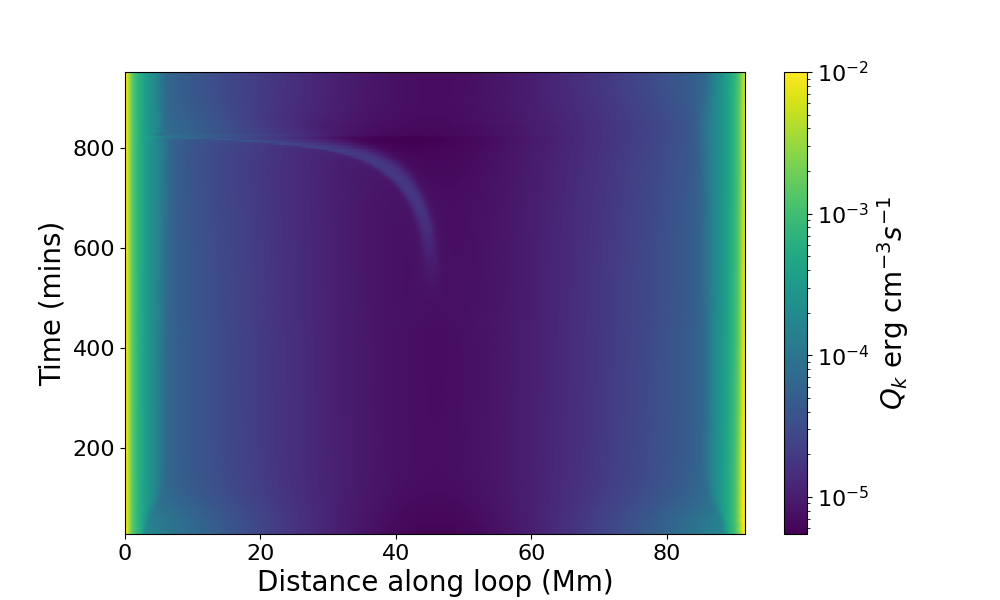}
    \end{minipage}\hfill
    \begin{minipage}[t]{0.2495\textwidth}
        \centering
        \includegraphics[width=\textwidth]{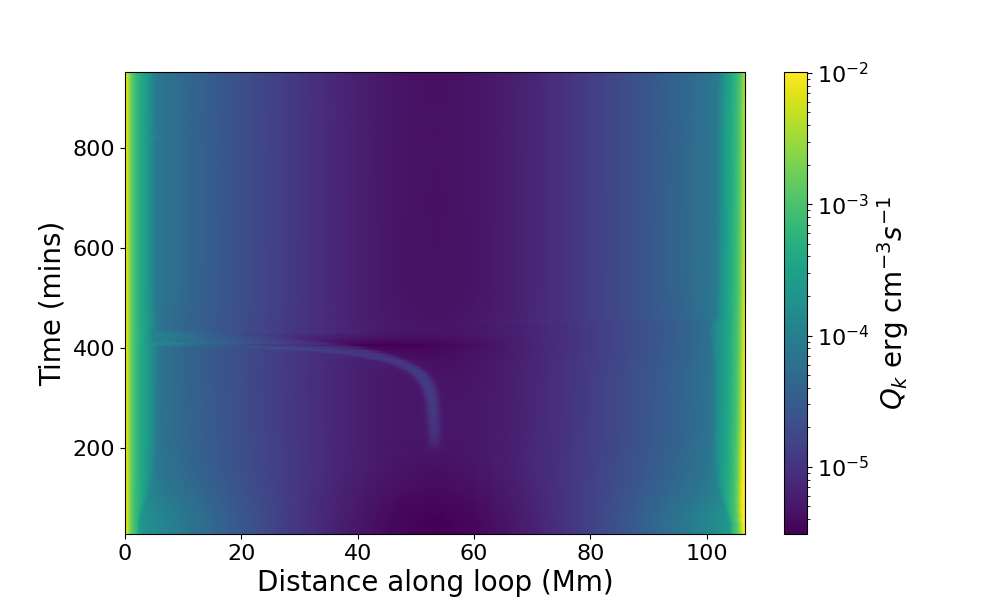}
    \end{minipage}\hfill
    \begin{minipage}[t]{0.2495\textwidth}
        \centering
        \includegraphics[width=\textwidth]{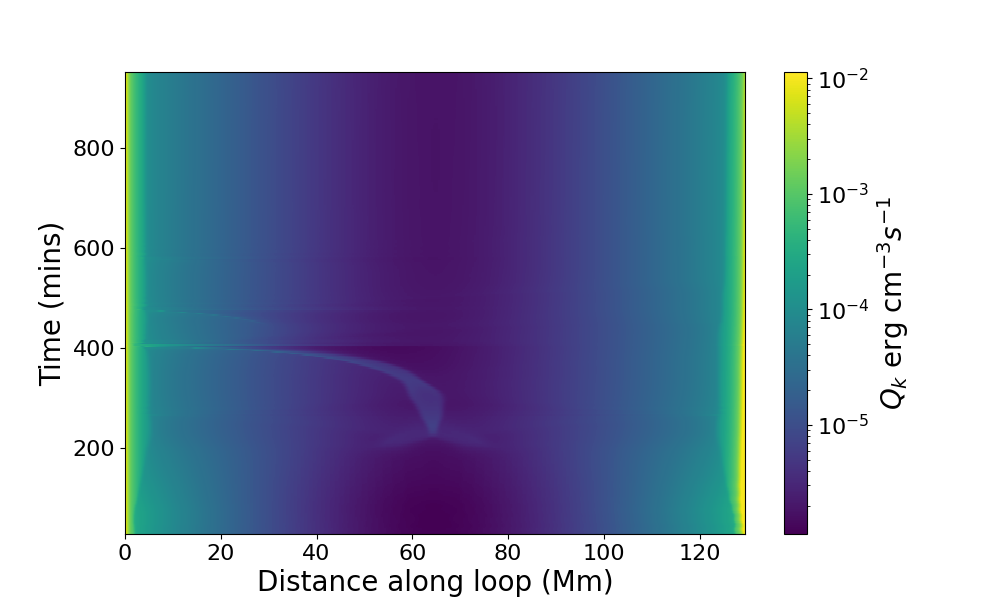}
    \end{minipage}\hfill
    \begin{minipage}[t]{0.2495\textwidth}
        \centering
        \includegraphics[width=\textwidth]{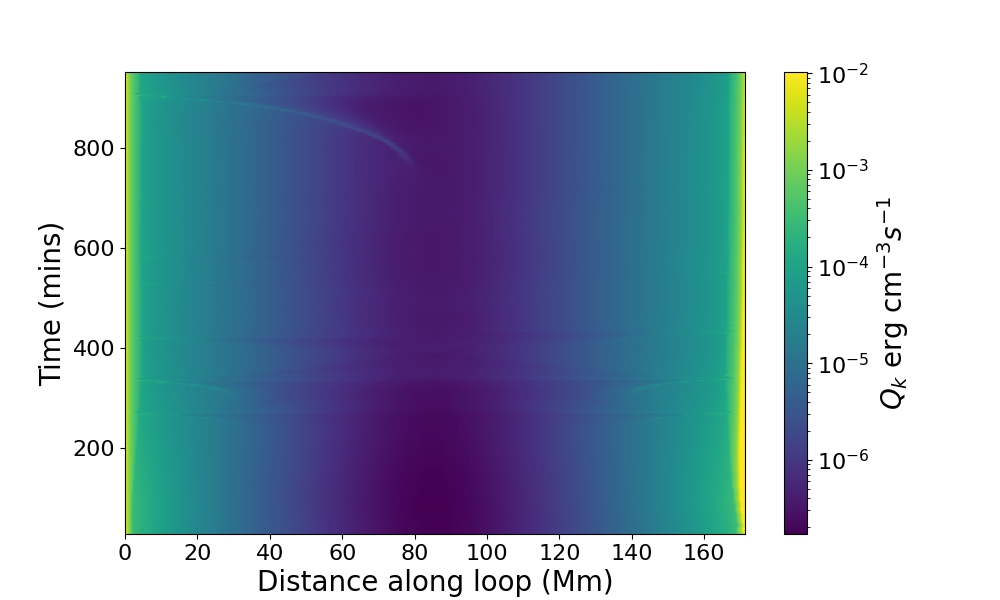}
    \end{minipage}

    \caption{The evolution of the density (top row) and kink wave heating rate (bottom row) along four magnetic field lines with footpoints located at $x = \pm$ 40, 42.5, 45, and 47.5 Mm, as a function of time.}
    \label{fig:field_line_plots}
\end{figure*}

Typical values for the oscillation amplitudes of kink waves in the corona can be estimated from our wave energy variables by adopting Equation (15) from \citet{TVD2014}, rewritten in terms of kink wave energy as given below
\begin{equation}\label{eq:energydensity}
w_\mathrm{obs} = \sqrt{\frac{2 W_{\mathrm{k}}(f \zeta + 1-f)}{f(1+\zeta)\rho_0}},
\end{equation}
where $w_\mathrm{obs}$ is the observed amplitude of the oscillation. We found the kink wave energy in our simulation corresponds to values of $w_\mathrm{obs}$ to be approximately $25\,\mathrm{km\,s^{-1}}$ at 35 Mm above the surface, which is comparable to the values observed by, e.g., \cite{Weberg2020}. 

\section{Discussion}\label{sec:Discussion}

Our results demonstrate that Alfv\'en and kink wave energy dissipation leaves thermodynamically distinct signatures in coronal loops when the injected wave energy is kept constant. The UAWSoM framework has revealed that Alfv\'en and kink waves differ not only in their ability to maintain a hot corona \citep{McMurdo2026}, but also in their capacity to generate the spatial and temporal heating stratification required for thermal instability. In particular, kink waves produce strongly structured heating due to their sensitivity to the geometry of the fine‑scale structures that make up coronal loops. As a result, neighbouring field lines can evolve in qualitatively different ways, with some remaining in quasi‑steady equilibrium for longer, while for others condensations form more rapidly. This behaviour is multidimensional and cannot be reproduced in traditional 1D coronal‑rain models, which impose a single heating profile along the loop and therefore cannot capture the multi-dimensional nature of coronal condensation formation.

We found that Alfv\'en waves do not trigger condensations in our setup due to their strong heating rate towards the loop apex. This clearly suppresses the development of the localised cooling regions required for thermal instability, in strong agreement with lower dimensional models of \cite{Antolin2010}. The absence of condensations in the Alfv\'en-wave-driven case therefore highlights a fundamental difference in how each wave mode perturbs the thermodynamic balance of the corona.

Our results also show that the onset of condensations in the kink-wave-driven case depends sensitively on the perpendicular correlation length. Reducing the correlation length by an order of magnitude concentrates the dissipation, steepens the heating stratification, and produces conditions favourable for thermal non‑equilibrium. This suggests that the fine‑scale structuring of coronal loop threads, in particular the radius, filling factors, and density contrast, play a decisive role in determining whether wave‑driven heating can trigger coronal rain. The velocity of the condensations in our simulations falls within the observed range of velocities of coronal rain, providing evidence that either mechanism, be it kink wave heating or localised heating from parameterised footpoint heating, or some combination of the two, could explain condensation formation. However, since the physical explanation of a local parameterised heating function is less justified than wave heating, this localised heating function could also be evidence of rapidly damped kink waves due to, e.g., fine-scale structuring of coronal loops below our chosen value of 100 km.

Analysis of the temporal evolution of density along multiple field lines in Figure \ref{fig:field_line_plots} (top panel) shows that condensations form differently and at different times on each magnetic field line. Although all strands receive the same injected wave energy at the base of the domain, this variation is a natural consequence of the expanding magnetic field and differing wave energy flux, and hence kink wave heating, occurring due to a transversely varying magnetic field, offering an explanation for the observed patchiness of coronal rain. 




\section{Summary and Conclusions}\label{sec:Summary and Conclusions}

Using the UAWSoM framework, we have shown that Alfv\'en and kink waves produce fundamentally different thermodynamic responses in coronal loops. For the same injected wave energy, Alfv\'en waves failed to generate condensations, whereas kink waves were able to trigger coronal rain when their dissipation became sufficiently localised. This establishes kink waves as a viable mechanism for driving thermal instability in the corona, provided that the fine‑scale structuring of the loop permits sufficiently short correlation lengths. Within the parameter range and symmetric 2.5D configuration considered here, Alfv\'en wave heating did not trigger condensations. The strong loop-apex heating found in the cases examined suppresses the localised cooling required for thermal instability. A broader parameter study, including fully 3D configurations, is required to determine whether this behaviour persists in more general models of coronal environments.

Our findings suggest there exist clear observable differences between Alfv\'en- and kink-wave-driven loops such as the presence or absence of condensations. While our model allows for the separation of the two wave modes, removing the effects of mode coupling and resonant absorption is not physical. In the real solar atmosphere, Alfv\'en-kink wave interactions may become important and it will certainly be a more complex task to distinguish between each wave's behaviour as in the case using our modelling framework. Once various physical parameters such as the radius of the flux tube, the width of the inhomogeneous layer, the density contrast, period and the velocity amplitude of the wave are estimated, it becomes possible to estimate the time scales of the various physical dissipation processes that affect each wave \citep{Goossens1992, Doorsselaere2020}. If the distinction is clear, retaining one physical dissipation process is justified. In the case that uniturbulence operates on the shortest timescale, we propose that the (non-)existence of cool structures may provide a pathway for mode‑specific or heating-mechanism-specific diagnostics of coronal heating.


Our model has chosen a single, idealised, magnetic field configuration, to aid comprehension of our simulation results. We note that the quadrupolar magnetic field topology chosen for this investigation has a relatively low expansion factor. In 1D models, this has been shown to favour condensation formation \citep{Mikic2013, Froment2018}, although it is not well known how realistic this result is with regards to higher dimensional models. The effect of time-varying magnetic fields is also not considered in the present investigation, although it will probably be an important effect, especially when condensations occur. As such this is left for future work. Additionally, our model's magnetic field is considered to be symmetric about $x=0$, and driven symmetrically by the various wave energy equations. The effect of asymmetry has been estimated by \cite{Klimchuk2019}, and seems to reduce the capacity for TNE to evolve. The choice of symmetry was intentional, and done in order to use this investigation to highlight the differing roles of each wave on the dynamics of a coronal loop in a simple model setup. As such we leave the topic of asymmetry for future work. In addition to these proposed future directions, we plan to extend these simulations to 3D geometries, incorporate more realistic magnetic topologies, such as the modelling of specific active regions, and compare synthetic observables with high‑resolution observations from DKIST, Solar Orbiter, Hinode and Hi-C. We predict that high-resolution simulations will make it possible to place constraints on a variety of currently unresolved coronal properties, such as the fine-scale structuring of the solar corona, that remain beyond the current spatial resolution of around 200 km as currently available using Hi-C. Despite \cite{Williams2020} finding that the most common strand widths were 500 km and are clearly resolved, it is our prediction that thinner structures need not dominate the domain for kink wave energy dissipation to have an important and potentially dominant impact on the thermal stability of the solar corona. \cite{Tamburri2025} used the Daniel K. Inouye Solar Telescope (DKIST) to resolve the widths of post reconnection flare loops and found values as small as 21 km, essentially the diffraction limit of DKIST. It is not known whether loops could be structured on even smaller scales, however, it is clear that higher resolution telescopes continue to detect smaller and smaller coherent structures. Although UAWSoM allows us to include the influence of unresolved fine-scale structuring parametrically through the density contrast, filling factor and characteristic radius, the present equilibrium does not geometrically resolve multiple coronal strands or prominence threads. We note that this fine scale structuring can significantly affect the dissipation rates of Alfv\'en \citep{Similon1989} and fast magnetoacoustic waves \citep{Murawski2001}. However, without considering the UAWSoM formulation, one would have to resort to extreme resolution simulations (beyond that of the aforementioned literature), which, beyond numerical expense, have their own limitations in terms of modelling turbulent wave dissipation. With this being said, an investigation that compares the phenomenological heating rate used in UAWSoM with the dissipation rates of kink waves in high-resolution simulations should be performed. 

In combination with observations of condensations and estimates of the available wave energy, our results may also enable constraints to be placed on various physical processes and parameters, including wave heating stratification. If the heating is assumed to be driven by kink wave dissipation, such constraints would provide valuable information on the correlation length and the physical quantities that determine it. This would allow us to determine the necessary resolution of future instruments that could finally resolve the fine-scale structuring of coronal loops on which waves dissipate fast enough/over short enough distances to be responsible for the dynamic behaviour observed in the corona. We tentatively predict that future observations should have sufficient spatial resolution to resolve coherent structures smaller than 100 km. Such a resolution would allow us to verify observationally whether coronal rain can be fully explained by kink wave heating across a range of short correlation lengths. In particular, it would enable us to determine whether the smallest correlation lengths that are required to produce the strongly stratified heating needed to form condensations exist.

\begin{acknowledgements}
MM and TVD received financial support from the Flemish Government under the long-term structural Methusalem funding program, project SOUL: Stellar evolution in full glory, grant METH/24/012 at KU Leuven. Furthermore, TVD was supported by a Senior Research Project (G088021N) of the FWO Vlaanderen. The research that led to these results was subsidised by the Belgian Federal Science Policy Office through the contract B2/223/P1/CLOSE-UP. It is also part of the DynaSun project and has thus received funding under the Horizon Europe programme of the European Union under grant agreement (no. 101131534). Dr. V.J.'s research was made possible by an appointment to the NASA Postdoctoral Program at the Goddard Space Flight Center, administered by Oak Ridge Associated Universities under contract with NASA. CF was supported by the Agence Nationale de la Recherche (ANR) for the CROSSWIND project under the grant ANR-24-CE31-2993, the Centre National d’Études Spatiales (CNES), France (ROR: https://ror.org/04h1h0y33), within the framework of the Solar-C mission, and the Action Thématique Soleil-Terre (ATST) of CNRS/INSU PN Astro, co-funded by CNES and CEA.. Views and opinions expressed are however those of the author(s) only and do not necessarily reflect those of the European Union and therefore the European Union cannot be held responsible for them.
    
\end{acknowledgements}

\bibliographystyle{aa}

\bibliography{bib.bib}

\end{document}